\documentclass[superscriptaddress,twocolumn,pre]{revtex4}
\usepackage{amsmath}
\usepackage{amsfonts}
\usepackage{adjustbox}
\usepackage{graphicx}
\usepackage{hyperref}
\usepackage{amsfonts}
\usepackage{bbm}
\usepackage{doi}
\usepackage{xcolor}

\begin{document}
\title{
Morse Theory and Meron Mediated Interactions Between Disclination Lines in Nematics}
\author{Joseph Pollard}
\email{joe.pollard@unsw.edu.au}
\affiliation{School of Physics, UNSW, Sydney, NSW 2052, Australia.}
\affiliation{EMBL Australia Node in Single Molecule Science, School of Medical Sciences, UNSW, Sydney, NSW 2052, Australia.}
\author{Richard G. Morris}
\email{r.g.morris@unsw.edu.au}
\affiliation{School of Physics, UNSW, Sydney, NSW 2052, Australia.}
\affiliation{EMBL Australia Node in Single Molecule Science, School of Medical Sciences, UNSW, Sydney, NSW 2052, Australia.}
\affiliation{ARC Centre of Excellence for the Mathematical Analysis of Cellular Systems, UNSW Node, Sydney, NSW 2052, Australia.}

\begin{abstract} 
    The topological understanding of nematic liquid crystals is traditionally centered on singularities, or defects, and their classification via homotopy theory. However, this approach has ultimately proved insufficient to properly capture a range of complex behaviours that have been reported in three dimensions. To address this, we argue that a finer understanding of topology is required, in which non-singular but non-trivial topological solitons---so-called merons---play a central role in mediating interactions between disclination lines. We present a comprehensive framework for capturing such behaviour that draws heavily on aspects of Morse theory; the key notion being that merons appear singular under projection onto a two-dimensional surface. This permits the use of singularity theory and dividing curves to characterise nematic textures via tomography, as well as an understanding of topological transitions via surgery theory. We use our ideas to understand and classify complex three-dimensional behaviours, such as the linking, rewiring and crossing of disclination lines, as well as to provide a new perspective on defect charge.
\end{abstract}
\maketitle

\section{Introduction}
Nematic ordering is a ubiquitous concept in physics, with applications ranging from molecular liquid crystals \cite{degennes_physics_2013} to morphogenesis in biological tissues \cite{saw_topological_2017, hoffmann_theory_2022}. The topological understanding of such materials is traditionally centered on singularities, or defects, which have proved fundamental to many aspects of their physics. As centres of strong elastic distortion, defects are the essential drivers of dynamics in both passive and active systems, and the behaviour of the system as a whole can often be approximated by the behaviour of the defects alone. For example, defects control the material's strength~\cite{taylor_mechanism_1934}, mediate elastic interactions~\cite{poulin_novel_1997, skarabot_two-dimensional_2007} and phase transitions~\cite{kosterlitz_ordering_1973, renn_abrikosov_1988}, as well as govern the coarsening dynamics of nematic order~\cite{chuang_coarsening_1993}. As a result, the physics of nematic defects has been studied extensively for many decades~\cite{kleman_classification_1977, kleman_relationship_1977, mermin_topological_1979, poenaru_topological_1979, poenaru_aspects_1981, kleman_defects_1989, kleman_points_1983, janich_topological_1987, kleman_disclinations_2008, copar_nematic_2011, copar_PRE_2011, copar_stability_2012, alexander_colloquium_2012, copar_quaternions_2013,  copar_topology_2014, martinez_mutually_2014, copar_knot_2015, afghah_visualising_2018, long_geometry_2021, schimming_singularity_2022, alexander_entanglements_2022, schimming_kinematics_2023, long_applications_2024}. In recent years the focus has shifted to active liquid crystals~\cite{marchetti_hydrodynamics_2013, ramaswamy_active_2017}, where the active flows cause $+1/2$ defects to self propel, inducing complex, defect-driven dynamics that have sparked a great deal of interest~\cite{pismen_dynamics_2013, giomi_defect_2013, giomi_defect_2014, decamp_orientational_2015, vromans_orientational_2016, tang_orientation_2017, urzay_multi-scale_2017, shankar_defect_2018, kumar_tunable_2018, tang_theory_2019, tan_topological_2019, copar_topology_2019, metselaar_topological_2019, carenza_rotation_2019, duclos_topological_2020, carenza_chaotic_2020, krajnik_spectral_2020, binysh_three-dimensional_2020, houston_defect_2022, alert_active_2022, houston_active_2023, kralj_defect_2023, kralj_chirality_2024, digregorio_coexistence_2024, wang_symmetry_2024, mitchell_maximally_2024}.

The classification of defects in terms of homotopy theory is a cornerstone of our understanding of these behaviours~\cite{mermin_topological_1979, alexander_colloquium_2012, degennes_physics_2013}. In addition to classifying the local structure of defects, homotopy theory provides a coarse means of describing local minimisers of the energy: two textures that correspond to distinct homotopy classes can only be deformed into one another by a process requiring the creation and annihilation of defects, which imposes an energy barrier between them. However, it has long been recognized that homotopy theory methods are insufficient in several respects. For example, they fail to fully describe materials with a spatially modulated ground state, such as smectics and cholesterics~\cite{poenaru_topological_1979, poenaru_aspects_1981}. Extensions of the homotopy theory using more advanced topological methods that analyse finer topological structure have proved invaluable in understanding the behaviour of these materials~\cite{chen_symmetry_2009, chen_topological_2012, kamien_topology_2016, machon_contact_2017, pollard_point_2019, machon_aspects_2019, eun_layering_2021, han_uniaxial_2022, pollard_contact_2023, pollard_escape_2024, severino_escape_2024}. Topological invariants arising from homotopy theory also fail to capture the complexity of defect textures in complex geometries, for example in colloidal systems~\cite{musevic_liquid_2017}, which has lead to the development of a more geometric theory~\cite{copar_nematic_2011, copar_PRE_2011, copar_stability_2012, copar_quaternions_2013, copar_topology_2014, copar_knot_2015}. More generally, characterising the complex spatio-temporal structures that emerge in both passive and active systems is a major challenge~\cite{martinez_mutually_2014, urzay_multi-scale_2017, copar_topology_2019, binysh_three-dimensional_2020, kralj_defect_2023, digregorio_coexistence_2024}, with simple topological invariants coming from homotopy theory being insufficient. 

Many of these failures manifest more prominently in three-dimensions than two. For example, in two dimensions the distinction between $+1/2$-winding and $-1/2$-winding defects is fundamental---you cannot convert one into the other. However, homotopy theory tells us that it {\it is} possible in three dimensions, and that there is no topological distinction between $+1/2$ and $-1/2$ winding around a disclination line~\cite{janich_topological_1987, alexander_colloquium_2012, alexander_entanglements_2022}. As a result, the topology of disclination lines has been downplayed due to a belief in its relative unimportance. However, we argue that this is an artefact of a homotopy-centric approach, which overlooks a notion that is very important to the underlying physics: in addition to topological defects, three-dimensional materials exhibit additional topological structures that, although non-singular, are nevertheless non-trivial. These are topological solitons, called merons or half-Skyrmions, which were originally introduced to describe the interactions resulting in quark confinement~\cite{callan_toward_1978}, and also occur in magentic systems~\cite{ezawa_compact_2011}. In liquid crystals, textures such as Skyrmion lattices and cholesteric blue phases---in which these solitons are the central features---are well studied both theoretically and in experiment~\cite{wright_crystalline_1989, fukuda_quasi-two-dimensional_2011, fukuda_ring_2011, kwok_cholesteric_2013, machon_global_2016, machon_umbilic_2016, nych_spontaneous_2017, tai_three-dimensional_2019, wu_hopfions_2022, pisljar_blue_2022}, but in other contexts they are often overlooked despite playing an important role in the dynamics of topological materials. In particular, while active Skyrmion lattices have been studied~\cite{metselaar_topological_2019}, the role of topological solitons in, for example, active turbulence, has not been appreciated. We argue that you cannot fully understand the behaviour of nematic textures without merons, which are shown here to mediate a variety of complex phenomena and interactions between defects, including changes in defect winding, as well as fundamental dynamical processes such as the rewiring and crossing of defect lines. 

In this work we provide a comprehensive and novel topological framework for the spatio-temporal analysis of nematic fields that both extends and enhances homotopy theory. This framework draws on techniques from the mathematical field of Morse theory~\cite{milnor_morse_1969} in order to provide a more detailed description of topological structure than homotopy theory alone can provide, and is applicable to both passive and active nematics, as well as ordered and elastic media more generally. Our framework unifies several disparate ideas from across liquid crystal physics, including the extended Volterra process~\cite{kleman_disclinations_2008}, the geometric theory of disclination lines as `ribbons'~\cite{copar_nematic_2011, copar_PRE_2011, copar_quaternions_2013, copar_topology_2014}, the homotopy theory description of disclination lines in both achiral and chiral nematics~\cite{janich_topological_1987, alexander_colloquium_2012, alexander_entanglements_2022, pollard_contact_2023, pollard_escape_2024}, and descriptions of the structure and orientation of disclination lines and solitons in terms of tensors that capture aspects of their geometry~\cite{machon_umbilic_2016, long_geometry_2021, schimming_singularity_2022, schimming_kinematics_2023}. 

We use this framework to give a topological classification of important changes of structure in a nematic, including dynamical crossing and rewiring processes, in terms of simple topological invariants which are easily computable in simulations and also from experimental data. A key insight of this work is the way that topological solitons mediate changes in defect structure and the interactions between defects, which has not previously been appreciated. We also revisit the problem of assigning a defect charge to a disclination line, where we argue that the conflation of Skyrmion charge and defect charge---reasonable in an orientable system---leads to confusion in the presence of disclination lines. We show that, in addition to global topological constraints, there are local topological constraints coming from Morse theory that have been overlooked. As a result, there are some errors and inconsistencies in previous works, which we correct. The tools we introduce provide a means of assigning defect charge that is both locally and globally consistent, and can be readily applied to textures of arbitrary complexity. 

The remainder of this paper is divided into two parts. The first part, encompassing Sections~\ref{sec:isotopy} to~\ref{sec:surgery}, introduces our new topological framework. In Section~\ref{sec:isotopy} we highlight the distinction between homotopy and isotopy for director fields, one aspect of which is the distinction between Volterra-type descriptions of crystalline defects and the homotopy theory description of liquid crystal defects. Isotopy invariants are finer than homotopy invariants, but still relate to energy barriers within the material. Changes in isotopy class can describe a process by which the material moves around within a homotopy class; while changes in homotopy class are mediated by defects, changes in isotopy class are mediated by merons. We highlight two key examples of the distinction: the conversion of a hedgehog point defect into a hyperbolic point defect, and the change in the winding of a disclination by the emission of a meron line. In Section~\ref{sec:tomography} we introduce the technique of tomography, in which a complex 3D director is studied through a family of surfaces that span the material. Tomography reduces the topological content of the director down to a finite set of features---defects and merons---along with the structural changes (or `degeneracies') in these features, several of which involve interactions between defects and merons. In Section~\ref{sec:surgery} we introduce the second part of our method, surgery theory. This provides a complementary perspective to tomography, and both are needed to fully analyse the topological aspects of dynamical processes such as rewirings. It also clarifies the notion of the extended Volterra process~\cite{kleman_disclinations_2008} by describing the homotopies which govern changes in the isotopy class of disclinations. 

The second part of the paper, Sections~\ref{sec:self_linking} to~\ref{sec:crossing}, applies this framework to several fundamental aspects of the behaviour of disclinations. In particular, our focus is on how dynamical processes involving disclinations, which include both changes in local structure via homotopies and also interactions between defects---rewirings and crossings---either introduce or remove structural degeneracies in the tomography. In Section~\ref{sec:self_linking} we explain how disclination lines are oriented and inherit a self-linking number related to the Burgers vector. This helps connect our work to previous studies that have focused on the geometry of disclination lines~\cite{copar_topology_2014, long_geometry_2021, schimming_kinematics_2023}, while also highlighting the fundamental topology that underlies their more geometric constructions. In Section~\ref{sec:defect_charge} we use surgery theory to describe how disclination lines of arbitrary knot type and with arbitrary local structure are endowed with a defect charge that is both locally and globally consistent: this is important for a proper understanding of how they interact with other disclination lines, and also with point defects. In Section~\ref{sec:rewiring} we give a topological classification of rewiring processes. While these have been studied previously, a complete topological classification is lacking. Indeed, we show there are several distinct types of rewiring, which has not previously been appreciated, and analyse the role merons play in mediating these processes through the formation of structural degeneracies in the tomography. Finally, in Section~\ref{sec:crossing} we describe the topological aspects of defect crossing processes. Throughout we illustrate our results with references to known textures, explain how our methodology applies and recapitulates known results as well as providing a novel perspective on these structures. 

We close in Section~\ref{sec:discussion} with a discussion of our work and prospects for future applications and developments.

\begin{figure*}[t]
\centering
\includegraphics[width=0.98\textwidth]{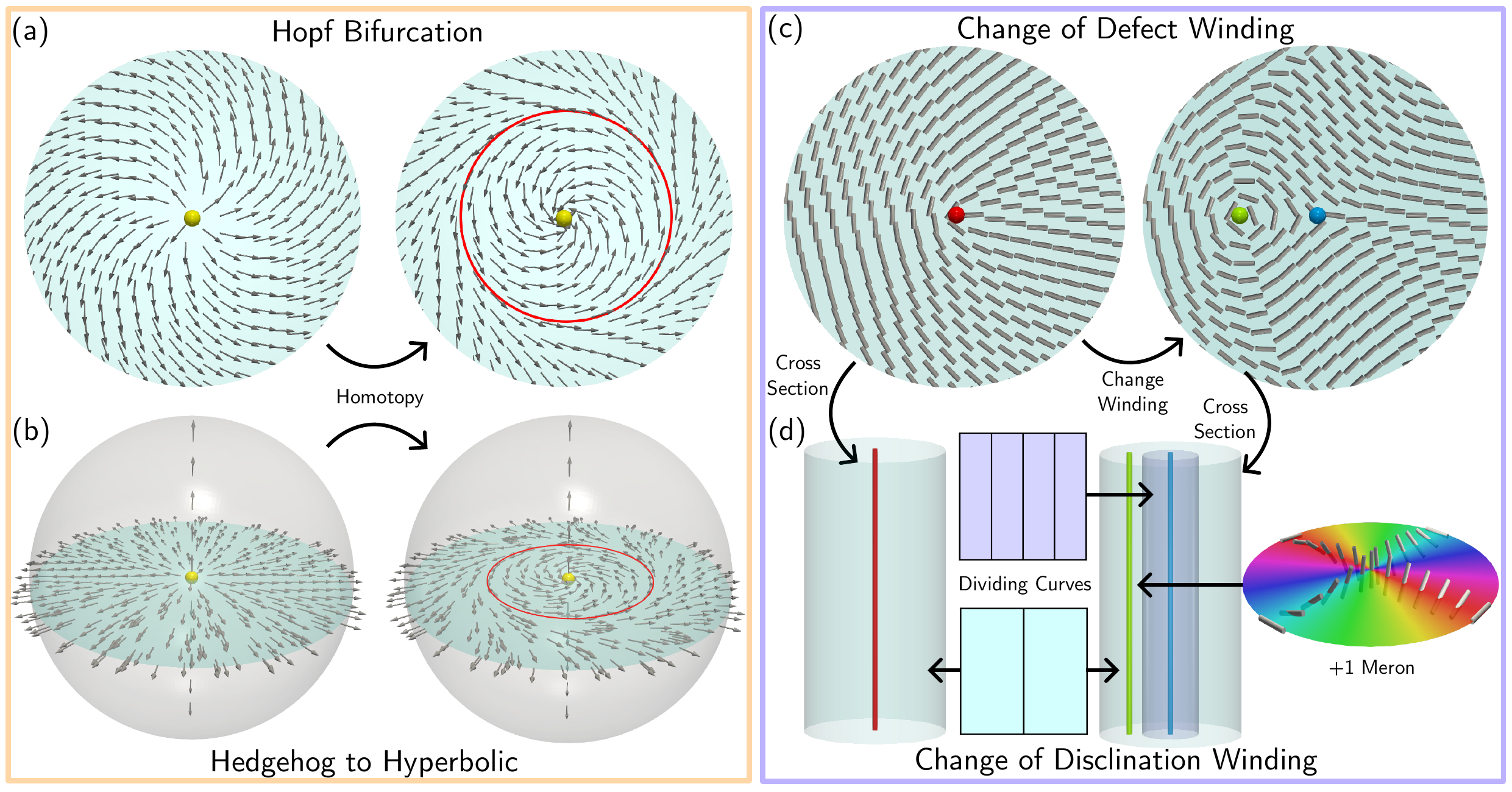}
\caption{Changes in a polar or nematic material that correspond to a homotopy that is not an isotopy can result in the birth of nonsingular soliton-like topological features that are analogous to dislocations, and which mediate dynamical behaviours. Understanding this finer topological constraint is fundamental to understanding wider phenomena that occur in 3D order materials, such as linking, rewiring and crossing of defect lines. The essentials can be captured by two simple examples. (a) A two-dimensional polar material containing an aster ($+1$ point defect, yellow) can undergo a Hopf bifuration, changing the stability of the singularity at the expense of expelling a limit cycle (red). (b) The analogous process in a 3D polar or nematic material is the conversion of a hedgehog defect to a hyperbolic defect. This occurs via a Hopf bifurcation in the director field on a cross-section. This process is a homotopy and preserves the charge, but as the structure of director field around a hedgehog is different from a hyperbolic defect we can distinguish them even when the orientation is removed. (c) We can change the winding of a point defect in a 2D nematic from $+1/2$ (red) to $-1/2$ (blue). The global constraint of defect charge conservation coming from the Poincar\'e--Hopf theorem implies we must emit a $+1$ singularity (green) in this process. (d) The analogue in 3D is the conversion of a $+1/2$-winding segment of a disclination into a $-1/2$-winding segment. Conservation of charge on a cross-section still applies: this process involves the emission of a $+1$ singular line. In 3D, integer-winding singular lines `escape', becoming nonsingular defects known as merons. The fact that $\pm 1/2$ disclinations belong to different isotopy classes is witnessed by the `dividing curve', introduced in Section~\ref{sec:tomography}.}
\label{fig1}
\end{figure*}

\section{The Distinction Between Homotopy and Isotopy Underpins Hidden Topological Structure in a Nematic}
\label{sec:isotopy}
A key aspect of this work is in identifying the role played by topological solitons, merons, in mediating changes of topological structure in a material. We argue that in order to properly account for merons, the traditional homotopy-centric approach to topological classification needs refinement. This involves two broad notions. The first is isotopy, a finer way of describing the material's evolution. Within a given homotopy class there are multiple distinct isotopy classes, and just as moving between different homotopy classes changes the defect-set---{\it i.e.}, the singularities---of a material, moving between different isotopy classes (within a given homotopy class) involves changes to merons. As homotopy classes correspond to large-scale energy barriers between textures, so do isotopy classes correspond to smaller energy barriers. 

The second important notion is the distinction between local and global homotopy classes. When a homotopy of the local structure of a defect is not an isotopy, then the homotopy must produce an additional feature in order to match on the rest of the texture. In the case of disclination lines this additional feature is a meron. 

We now review the concepts of homotopies and isotopies. We describe the distinction using two key examples of relevance to real systems; we will return to these examples repeatedly throughout the remainder of the paper, as they illustrate the main notions we wish to convey.

\subsection{Homotopies vs. Isotopies}
Throughout this work, the state of the nematic material is captured by a unit vector ${\bf n}$ called the director, defined on the complement of some set of defects. The topological theory of nematic liquid crystals is based on homotopies, smooth paths ${\bf n}_t$ of director fields, which typically move the defect set but not create, destroy, merge, or cross defects. Homotopy theory therefore concerns the computation and interpretation of quantities which are left invariant by homotopies, such as the total defect charge. More broadly, it has been used to classify the local structure of both point and line defects defects~\cite{mermin_topological_1979, janich_topological_1987, poenaru_crossing_1977, alexander_colloquium_2012, alexander_entanglements_2022}, and it classifies textures on the complement of the defects in terms of an Euler class (Skyrmion charge)~\cite{machon_umbilic_2016, machon_global_2016, bott_differential_1982} and Hopf invariant~\cite{chen_generating_2013, wu_hopfions_2022}. 

Homotopies are the natural model for the evolution of a nematic material, where the `sticks' may move independently of the material background. In other kinds of elastic media, for example crystals, elastic solids, and fluids, the state of the material at time $t$ is instead captured by a diffeomorphism $\psi_t$ which encodes the deformation away from some reference configuration; the vector field ${\bf u}_t = \partial_t \psi_t \circ \psi^{-1}_t$ is the displacement field, whose symmetric gradient is the strain tensor~\cite{landau_theory_2009}. The evolution of an elastic solid away from its initial state is captured not by a homotopy, but rather an isotopy, which is a smooth path of diffeomorphisms $\psi_t$ with $\psi_0$ the identity map. 

We may consider isotopies in a director field as well: A director field ${\bf n}_t$ is isotopic to another director field ${\bf n}_0$ if there exists an isotopy $\psi_t$ with $\psi^*_t {\bf n}_t =  \lambda_t {\bf n}_0$ for some scalar function $\lambda_t$. Informally, an isotopy between directors means they are related by a coordinate transformation of the material domain. Isotopies must therefore map integral curves of the director (lines to which the director is tangent) to other integral curves. A consequence of this is that isotopies may not create or destroy merons, while homotopies may create or annihilate these features freely so long as the Euler class and Hopf invariant are conserved. Thus, the role of isotopies in the study of nematics is to pick out finer changes in the topology structure of the material which the homotopy theory is insensitive to. 

The distinction between isotopy and homotopy has been discussed previously by Kleman in the context of the Volterra process for nematic defects~\cite{kleman_classification_1977, kleman_relationship_1977, kleman_points_1983, kleman_disclinations_2008}. Defects (dislocations) in a crystal are described by the Volterra process~\cite{volterra_sur_1907}, in which we imagine cutting a tube out from the material, slicing the tube open, displacing the sides of the cut relative to one another via some rotation and translation, and then replacing the tube. The transformation we use to perform the displacement provides a topological label for the type of dislocation which is an isotopy, but not a homotopy, invariant. As Kleman noted~\cite{kleman_classification_1977, kleman_relationship_1977, kleman_points_1983, kleman_disclinations_2008}, this process has to be refined in a liquid crystal because of the possibility of performing homotopies. Our framework helps to further elucidate this distinction, as we explain in Section \ref{sec:surgery}.

\subsection{Example 1: The Conversion of a Hedgehog Point Defect into a Hyperbolic Point Defect}
Up to homotopy, point defects are classified purely by their integer charge, generically $\pm 1$. This is obtained by surrounding the point with a sphere $S$. The director field can be oriented across $S$, and it then determines a map $S \to S^2$ whose degree (winding) is the defect charge~\cite{mermin_topological_1979, alexander_colloquium_2012}. However, point defects can have geometrically distinct local structures: they may be radial hedgehogs, where the director points out everywhere from the singular point, or they may have a hyperbolic structure. The distinction is given by an isotopy invariant called the Morse index, which we describe properly in Section~\ref{sec:tomography}.The Morse index is not a homotopy invariant, so we may freely convert between a hedgehog and hyperbolic structure by homotoping the director in a local neighbourhood of the defect. This process has be observed experimentally in cholesteric droplets with radial anchoring~\cite{posnjak_hidden_2017}, where the material's chirality induces the conversion~\cite{pollard_point_2019}. 

Because this homotopy is only a local change there are nontrivial questions about how it extends globally. First, let us briefly examine the analogous situation in 2D, Fig.~\ref{fig1}(a). In a 2D polar material a $+1$ winding singularity can change its stability, with the polar field going from pointing out from the defect to pointing in, by emitting a limit cycle; this can also occur to a $+1$ singularity in a nematic field, whether it is oriented or not. This is the Hopf bifurcation, which we illustrate in Fig.~\ref{fig1}(a). The limit cycle, shown in red, can (informally) be regarded as a nonsingular defect, a kind of topological soliton. It cannot be freely removed by local rearrangements of the director, and can only be destroyed by collapsing it into a singularity or merging it with another limit cycle. 

The 3D version of the Hopf bifurcation, shown in Fig.~\ref{fig1}(b), is how a radial hedgehog point defect converts into a hyperbolic point defect: an a cross-section, the process is exactly a Hopf bifurcation. Such `twisted hedgehogs' are known to occur in chromonic liquid crystals, as well as materials whose splay elastic constant is significantly greater than their twist elastic constant~\cite{lavrentovich_phase_1986, paparini_spiralling_2023, ciuchi_inversion_2024}. 

Imagine removing a small ball of material containing a hedgehog point defect and trying to glue a hyperbolic defect in its place. Because they are homotopic it is possible to do this without inducing additional defects on the boundary of the gluing region, so long as the defects carry the same charge. However, we must introduce a limit cycle as in Fig.~\ref{fig1}(b) that witnesses the fact that the hedgehog belongs to a different isotopy class to the hyperbolic defect---the presence of this feature is a strict topological constraint.

\subsection{Example 2: The Change in the Winding of a Disclination Line}
An even more important example of this topological constraint occurs for disclination lines. Unlike the situation for a 2D point defect, the winding of the director around the disclination line, generically either $\pm 1/2$, is not a homotopy invariant, and it can freely change along the line and also over time~\cite{mermin_topological_1979, alexander_colloquium_2012}. However, there is a subtlety here. 

Consider the conversion of a $+1/2$ defect into a $-1/2$ defect in a 2D system Fig.~\ref{fig1}(c). The constraint of conserving topological charge---the Poincar\'e--Hopf theorem---illustrates that this requires the production of an additional $+1$ defect. This process is therefore not a homotopy by our definition, because it changes the defect set. Extending the 2D example into 3D, Fig.~\ref{fig1}(d), shows that a $+1/2$ profile disclination line must still emit a $+1$ winding singular line to convert to $-1/2$ winding. In a 3D material integer winding singular lines are unstable and are well-known to `escape into the third dimension' to produce a meron line~\cite{meyer_existence_1973, cladis_non-singular_1972, pollard_escape_2024, lequeux_helicoidal_1988, murray_decomposition_2017}. Therefore, this process of changing the winding does not involve a change in the defect set, and occurs via a homotopy. However, the constraints coming from the Poincar\'e--Hopf theorem also apply to a conservation of winding number for 3D disclinations, albeit in a much subtler way that we will elucidate in Sections~\ref{sec:tomography} and~\ref{sec:surgery}.

Consider now removing a region of material containing a $+1/2$ disclination, indicated by the blue tube in Fig.~\ref{fig1}(d)(i). Suppose then we glue in a $-1/2$-winding disclination, as in the purple tube in Fig.~\ref{fig1}(d)(ii). Because these textures are homotopic this can be done without introducing additional defects, but because they are not isotopic a meron must appear. As with the point defects, the presence of this feature is required by a strong topological constraint which cannot be violated. Showing this, and describing the consequences for structural changes in the material, is the goal of the following Sections~\ref{sec:tomography} and~\ref{sec:surgery}.

\begin{figure*}[t]
\centering
\includegraphics[width=0.98\textwidth]{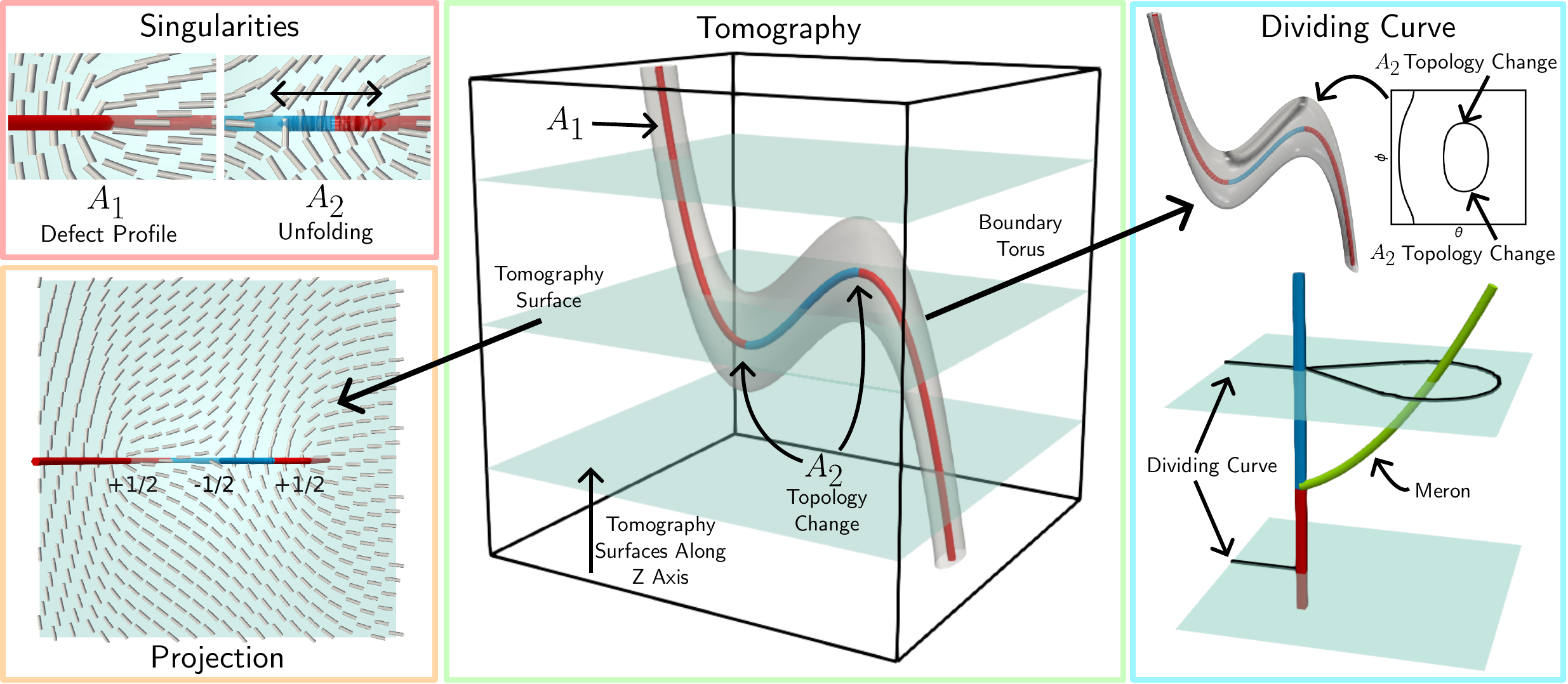}
\caption{A tomographic analysis of a director field identifies and characterises not only disclinations, which are singular, but also merons, which are non-singular. We consider sliding a surface (pale blue) across a 3D director field that contains a disclination line. The projection of the director field encodes the topology of the director along that surface. As we slide the surface across the material, we obtain a family of projections that determines the entire topological structure of the director field. Crucially, the projection has singularities where it intersects not only disclinations but also merons, which are nonsingular in 3D. We colour the disclination according to whether or the singularity it induces in the projection has winding $+1/2$ (red) or $-1/2$ (blue). The isotopy class changes as the tomography traverses surfaces on which the number and type of singularities in the projection change. The changes in topological structure of the disclination line in the central panel can be seen in the dividing curve on a tube surrounding the disclination (top right). The isotopy class of the dividing curve associated to the projections into the tomography surfaces also changes as the tomography traverses surfaces over across which the isotopy class of the director changes. This is shown schematically on the bottom right, for a scenario in which a disclination changes its winding from $+1/2$ to $-1/2$ by the emission of a meron line (green). The dividing curve (black) grows an additional loop, witnessing the change. }
\label{fig2}
\end{figure*}

\section{Tomographic Analysis Reveals the Topological Structure of Nematic Textures}
\label{sec:tomography}
We now describe a tomographic approach to understanding 3D textures. This is designed to capture merons by analysing 2D planes in which their projected topology appears singular, and the constraints coming from the Poincar\'e--Hopf theorem can be seen to apply. The result of this analysis is to break the material up into regions in which the structure can be described purely by an isotopy, separated by surfaces in which the isotopy class changes due to the birth or death of a topological feature. 

In brief, our process is this. We choose a family of surfaces that spans the material, say surfaces of constant $z$-coordinate. The structure of the director along each surface is captured by its projection and the topology is unchanged provided the topology of the projection does not change. Changes occur suddenly in jumps, and signify fundamental topological changes in the material which we characterise using singularity theory. Between the jumps, one may apply a diffeomorphism to bring the director into a form in which it is independent of the $z$-coordinate---these can be viewed as fundamental building blocks of the structure, and the structure of the director can then be described coarsely by these blocks and the transitions between them. This approach is close to current experimental methods for the analysis and reconstruction of director fields using flourescent confocal polarisation microscopy (FCPM)~\cite{smalyukh_three-dimensional_2001, posnjak_points_2016, posnjak_topological_2018}. 

We summarise our approach graphically in Fig.~\ref{fig2}. There are four parts to the theory: (A) the analysis of the structure of the director near to a surface and the detection of topological features, the disclinations and the merons, (B) how this changes as we move the surface through the material, (C) a label, derived from singularity theory, that encodes local structure of disclinations and merons as seen by the projections, and (D) a classification of changes in this local structure, again in terms of singularity theory.

\subsection{Projections and the Dividing Curve Determine the Local Topology of the Director Along a Surface}
Let $S$ be an oriented surface inside the material. Along $S$ the director defines a line field, the projection ${\bf n}^\perp$ of the director into the surface. As ${\bf n}$ is a unit vector the projection completely determines the director along $S$, a fact which is used to reconstruct experimental directors from light microscopy data~\cite{smalyukh_three-dimensional_2001, posnjak_topological_2018}. The projection has singularities in two places: $\pm 1/2$ singularities at points where the surface intersects disclination lines, and $+1$ singularities where the director is orthogonal to $S$. 

The director can be homotoped so that it is tangent to $S$ around all disclinations, and indeed everywhere on the surface except on a set of disks $D_j$ around the points $p_j$ where the director is orthogonal to the surface~\cite{machon_global_2016, pollard_escape_2024}. These singularities represent the local obstruction to homotoping the director to make it tangent to $S$---they are merons, and in the absence of disclinations they define a topological invariant of the director called the Euler class. 

Suppose there are no disclinations intersecting $S$. Then the director can be oriented in a neighbourhood of $S$. Let $p_j$ be the points where the director is orthogonal to $S$, which have winding $w_j$ and orientation $s_j = \pm 1$ according to whether the director points out of, or into, the surface---that is, whether the singular line has escaped `up' or `down'. On a disk surrounding each meron point the director determines a class in the relative homotopy group $\pi_2(\mathbb{RP}^2, \mathbb{RP}^1) \cong \mathbb{Z} \times \mathbb{Z}$~\cite{machon_global_2016, pollard_escape_2024}. For $s_j = +1$ this class is $(w_j, 0)$ and for $s_j=-1$ it is $(0, -w_j)$. The value of the Euler class on $S$, equivalently the Skyrmion charge $Q$, is equal to~\cite{pollard_contact_2023, pollard_escape_2024, copar_topological_2012},
\begin{equation} \label{eq:euler_class}
  2Q = \sum_j s_j w_j. 
\end{equation}
The topology of the director can be encoded in an even simpler object than the projection. Consider the set of points $\Gamma$ where the director is tangent to $S$. After perhaps performing a small homotopy of the director, $\Gamma$ is a collection of disjoint curves referred to as the `dividing curve'~\cite{geiges_introduction_2008, pollard_contact_2023, pollard_escape_2024}. Generically the dividing curve consists of a finite number of disjoint, one dimensional curves; if this is not the case, then we may perturb the surface slightly to make it so. 

When $S$ does not intersect disclinations then the components of the dividing curves will either be closed loops, or will have endpoints on the boundary of $S$ in the case where $S$ has a boundary. When $S$ intersects disclinations there will also be components of the dividing curve with endpoints on the disclinations---a $+1/2$ singularity will generically have one component of the dividing curve terminate on it, a $-1/2$ singularity will have three. A $+1$ meron is encircled by a closed loop. If the director can be oriented along $S$, the dividing curves separate regions where the director points into $S$ (the set $S^-$) from regions where the director points out from $S$ (the set $S^+$), which is the origin of the terminology. This leads to an alternative formula for the Skyrmion charge~\cite{pollard_contact_2023, pollard_escape_2024, copar_topological_2012}, 
\begin{equation} \label{eq:euler_class2}
  2Q = \chi(S^+)-\chi(S^-),
\end{equation}
where $\chi$ denotes the Euler characteristic. Note that if $S$ is a sphere, then the value of $Q$ computed by \eqref{eq:euler_class2} is exactly the defect charge contained within that sphere---in particular, this implies that the sphere is pierced by some number of merons, and indeed that merons mediate interactions between point defects~\cite{pollard_point_2019}. When $S$ is not a sphere, a nonzero $Q$ does not imply there are any defects in the material---the simplest counterexample being a surface $S$ cutting across a Skyrmion lattice. 

We see how the 2D Poincar\'e--Hopf theorem manifests itself here: it applies directly to the projection ${\bf n}_\perp$ on $S$, requiring that the sum of the windings of the singularities be equal to the Euler class of $S$, modulo the boundary condition. For example, if $S$ is a cross-sectional disk of a sphere or cylinder with radial or planar anchoring, the winding of the singularities of ${\bf n}_\perp$ must add up to $+1$. Locally, this constraint applies to a disk $D$ around a singularity too: if a homotopy localised to the disk changes the winding of the singularity, say from $+1/2$ to $-1/2$ as illustrated in Fig.~\ref{fig1}(c), then it must produce a $+1$ singularity in the projection which witnesses the meron emission. 

A key aspect of the theory of dividing curves is that they allow for `gluing' constructions. Suppose we separate a material domain into two pieces $M_0, M_1$ with a surface $S$ between them. Suppose we define a director ${\bf n}_0$ on $M_0$ and ${\bf n}_1$ on $M_1$. It is possible to glue these two pieces together along the common boundary $S$ without introducing any form of defect or soliton if and only if ${\bf n}_0, {\bf n}_1$ induce the same dividing curve on $S$~\cite{geiges_introduction_2008}. In mathematics these kinds of constructions, which are closely related to Morse theory, fall under the purview of `surgery theory'~\cite{kirby_topology_1989, gompf_4-manifolds_1999}, which we describe in more detail in Section \ref{sec:surgery} below. For now, we observe that the dividing curve already shows that the void left by removing a $+1/2$ defect cannot be filled by just a $-1/2$ defect---the dividing curves don't line up, as illustrated in Fig.~\ref{fig1}(d)---and the gluing process must be mediated by a meron. 

Finally, for completeness we explain how projections and dividing curves can be defined in the Q-tensor framework. Let $Q$ be the tensor defining the nematic order. In a neighbourhood of a point on a surface $S$ we can find a system of local coordinates $(x,y,z)$ so that the $z$-axis is always normal to the surface. We have an operation $J = {\bf e}_z \times$ that effects an anticlockwise rotation. The space of traceless, symmetric $3 \times 3$ tensors along $S$ is spanned by,
\begin{equation} \label{eq:Qtensor_basis}
    \begin{aligned}
    D_1 &= e_x \otimes e_x - e_y \otimes e_y, \\
    D_2 &= JD_1 = e_x \otimes e_y + e_y \otimes e_x, \\
    D_3 &= -e_x \otimes e_x + e_z \otimes e_z, \\
    F_1 &= e_x \otimes e_z + e_z \otimes e_x, \\
    F_2 &= e_y \otimes e_z + e_z \otimes e_y.
    \end{aligned}
\end{equation}
These matrices are all orthogonal with respect to the Euclidean inner product. It is clear that the matrices $D_1, D_2$ span the subspace of 2D traceless and symmetric tensors, which therefore define a line field on $S$. Using this decomposition we may then define an orthogonal projection that sends a 3D tensor $Q$ to a 2D tensor $Q^\perp$, which acts by dropping the terms involving $D_3, F_1, F_2$. Note that, while a director field can be reconstructed from the projection, the Q-tensor generally cannot be. This is because it encodes an entire frame, and we require more information to reconstruct a frame than just the projection~\cite{pollard_intrinsic_2021, da_silva_moving_2021}. 

We have written this in terms of a local coordinate basis, but it is clear this construction generalises to give a projection operator $\Pi_S$ that acts on $Q$ in the 3D space to produce the 2D tensor $Q_\perp$ over the entire surface $S$, and an operator $J$ that affects a rotation about the unit normal to $S$. Zeros of $Q^\perp$ correspond both to disclinations and merons, just like singularities in the projection. The line field ${\bf n}^\perp$ corresponding to the positive eigenvalue of $Q^\perp$ defines the projection.

\subsection{Tomography Reduces a Complex 3D Director to a Family of 2D Directors That Encodes the Global Topology}
To capture the full structure of a 3D director we choose a family of surfaces $S$ which span the material. A natural choice is to take surfaces $S_z$ of constant $z$ in a Cartesian coordinate frame $x,y,z$, but we will also at times consider a family of torus tubes $S_r$ around a disclination line $K$ of varying radius $r$; these will generally only span a neighbourhood of the disclination, not the entire space. A natural physical justification for a particular choice of surfaces is that they are tangent to the far-field director, which in our case we take to be aligned with the $y$-direction. 

The family $S_z$ is indicated by the blue surfaces in Fig.~\ref{fig2}. On each surface we consider the projection ${\bf n}^\perp_z$ of the director ${\bf n}$ into the surface, which defines a family of 2D line fields parameterised by the coordinate $z$, as in Fig.~\ref{fig2}. If we have dynamics then we have a family ${\bf n}_t$ of director fields evolving in time, which leads to a two-parameter family of 2D line fields ${\bf n}^\perp_{z,t}$ on our family of surfaces that fully captures both the spatial and temporal structure of the director. We refer to this as the `tomography' of the director. 

The tomography defines a family of sets of points $\Gamma_{z,t}$ where the director is tangent to the surface $S_z$ at time $t$. As we have remarked, for a generic choice of surface $\Gamma_{z,t}$ will be a dividing curve, consisting of a finite number of disjoint closed loops and lines. However, we have a parameterised family of surfaces: the generic condition is now that each $\Gamma_{z,t}$ is a dividing curve, except for specific isolated values of $z$ and $t$ where a component of the dividing curve undergoes a birth/death, merging, or crossing event. Between parameter values where these events take place, the director can be isotoped to a quasi-2D director field that is independent of $z$ and $t$. Thus, the essential topological features of the time-dependent director ${\bf n}_t$ boil down to the behaviour at specific values of $z$ and $t$ where these events occur.

Jumps in the isotopy class correspond to crossing, either spatially or temporally, a feature such as a disclination, point defect, or meron line. Tomography shows a duality between two processes: moving the surface spatially through the material, we cross through a meron line; keeping the surface fixed but having the director change over time, a meron is emitted. When the dividing curves on two surfaces are not related by an isotopy there has to be a defect between them---this defect may either be another disclination, or else a meron. Equivalently, if we view the transition as occurring in time rather than in space, the same argument shows that a defect must be created in the transition process. This duality between spatial and temporal changes provides a novel perspective on the structure and dynamics of a material.

At any point where the dividing curve changes there will also be a change in the structure of the singularities in the projection ${\bf n}_{z,t}^\perp$. In what follows we focus on the singular set rather than the dividing curve---although we continue to make use of the idea of the dividing curve in other contexts---and describe the types of changes that may occur presently. 

It is possible that some changes in the singular set (or dividing curve) are artefacts of the family of surfaces chosen, and not topological changes in the material. Whether or not this is the case can be determined by examining the dividing curves on tubes surrounding the disclinations, which we discuss more in Section \ref{sec:surgery}. For example, consider the structure shown in Fig.~\ref{fig2}. This disclination `bends back' on itself. We might imagine that we can simply `straighten it out' by a homotopy. However, this structure is \textit{not} homotopic to a straight disclination line which does not `bend back' on itself, which can be seen from the dividing curve on the encircling torus: the only way this curve can deform to a straight line is by having the central $-1/2$-winding section merge with one of the $+1/2$ sections.

\begin{figure*}[t]
\centering
\includegraphics[width=0.98\textwidth]{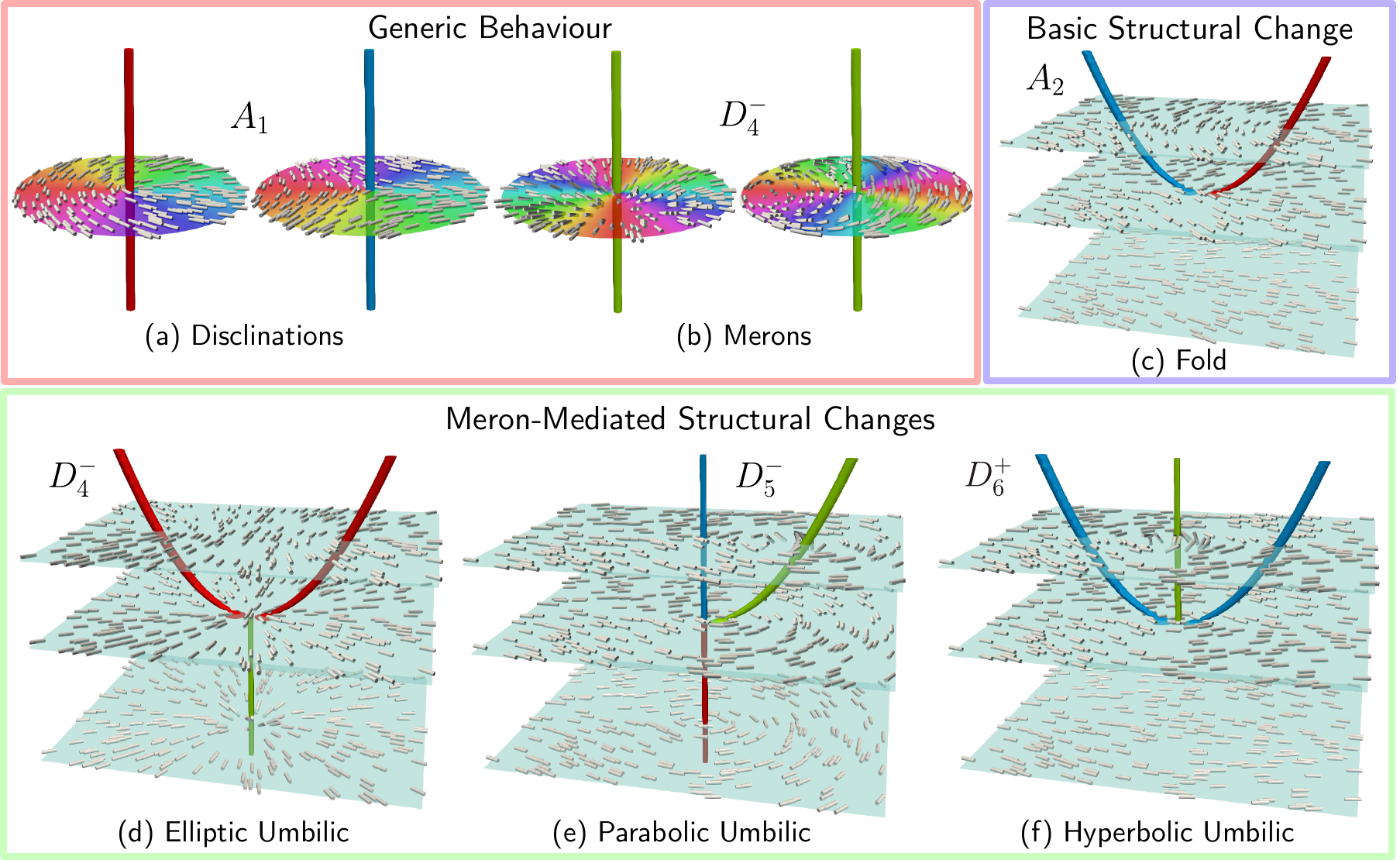}
\caption{Singularity theory provides a classification for possible structural changes in the topology of a director, and identifies the role of merons in mediating these changes. At a generic point, a singularity in the projection corresponds either to (a) a disclination of winding $+1/2$ (red) or $-1/2$ (blue), or (b) a meron line with winding $\pm 1$; illustrated are two cases of $+1$ winding, radial (left) and rotational (right). In terms of the associated vector field ${\bf q}$ defined in the main text disclinations and merons belong respectively to the singularity classes $A_1$ and $D_4^-$. The cross-section is coloured according to the angle between the $x$ and $y$ components of the director. (c-f) At isolated nongeneric points the director corresponds to a different singularity class which describes a change in structure, and away from these points the structure is given by an unfolding of the singularity. (c) The most simple structural change involves the meeting of disclination strands of opposite winding $+1/2$ (red) and $-1/2$ (blue), which is given by the $A_2$ singularity. There are then three structural changes mediated by merons. (d) The meeting of strands of the same winding is given by the $D_4^-$ singularity---when the strands merge there is a $+1$ singularity, which will generically be escaped to form a meron (green). (e) The emission of a meron line which remains tethered to the disclination is described by the $D_5^-$ singularity. (f) A more complex structure involving a cusp point in a disclination which meets a meron line, given by the $D_6^+$ singularity.}
\label{fig3}
\end{figure*}

\subsection{Singularity Theory Determines The Local Structure of Disclinations and Merons}
A singularity is a zero of a vector or line field. We consider only point singularities here, which may be either point defects in the director (see Ref.~\cite{pollard_point_2019}) or the singularities in the projection of the director into a surface $S$. Line singularities will be described via tomography, in terms of the point singularities in the family of 2D projections. 

As we discussed in Section \ref{sec:isotopy}, a point singularity in either a vector field or a line field has a single homotopy invariant, the defect charge~\cite{mermin_topological_1979, alexander_colloquium_2012}. Besides this homotopy invariant, point singularities also have diffeomorphism invariants that provide a more refined description of their local structure. Around a zero $p$ of a vector field ${\bf u}$ we can consider local diffeomorphisms $\psi$ that fix the origin and map ${\bf u}$ to $\psi^* {\bf u}$. A `singularity' is an equivalence class of such maps~\cite{arnold_singularity_1998}. 

Basic orientable singularities are Morse type~\cite{milnor_morse_1969}. These are associated to vector fields ${\bf u} = \nabla f$, where $f$ is a `Morse function'~\cite{milnor_morse_1969, arnold_singularity_1998, pollard_point_2019} of the form,
\begin{equation} \label{eq:morse_singularity}
    f = \pm x^2 \pm y^2 \pm z^2.
\end{equation}
The number of minus signs in this function is the `Morse Index' (MI) of the defect, which is invariant under diffeomorphisms. These models define the local structure of generic 3D point defects~\cite{pollard_point_2019}: those with MI 0 or 2 have defect charge $+1$, and those with MI 1 or 3 have defect charge $-1$. MI 0 and 3 defects have a `hedgehog' structure, with the director field pointing radially out from the defect everywhere, while those with MI 1 and 2 have a geometrically distinct `hyperbolic structure'. In 2D, we drop the $z^2$ term. The Morse index therefore varies from 0 to 2: singularities with MI 0 or 2 are radial and have winding $+1$, and those with MI 1 are hyperbolic and have winding $-1$. 

Morse singular points are generic in the sense that any singular point can be made Morse by a small perturbation. However, there are two subtleties to this. Firstly, the conversion may require breaking the singularity apart into multiple other singularities; secondly, if we have a one-parameter family of vector fields, for example in a tomography of a director or along a disclination line, it is {\it not} the case that the singularities can be made Morse for {\it every} value of this parameter. That is, whilst a perturbation can makes the singularities of a given, single plane Morse, it necessarily alters neighbouring planes, since the texture must be smooth, effectively just moving the obstruction elsewhere. 

In this context, singularity theory gives a framework for describing the local structure of singularities and the ways they change and break apart as we slide our surface through the material. The classification of Arnold \textit{et al.}~\cite{arnold_singularity_1998} describes all possible singularities in vector fields that are of the form $\nabla f$, for $f$ a polynomial function---different local structures that are not gradients of functions are of the form $R_\alpha \nabla f$, where $R_\alpha$ is a rotation about the surface normal by an angle $\alpha$. The classification scheme also enumerates the possible `unfoldings' of the singularities, which describe the ways in which they split apart into singularities of lower complexity (e.g., a number of Morse-type singularities with varying Morse Indices) as a parameter is varied. 

To describe changes in the tomography of a liquid crystal we need the classification of singularities and their unfoldings in \textit{line fields} rather than \textit{vector fields}, which are not explicitly covered by Arnold \textit{et al.}~\cite{arnold_singularity_1998}. However, in 2D this classification is simple to obtain from the vector field classification using the decomposition of the space of Q-tensors in Eq.~\eqref{eq:Qtensor_basis}. We see that there is an isomorphism between $Q$-tensors and vector fields in 2D, obtained by sending the tensor $Q = q_x D_1 + q_y D_2$ to the vector field ${\bf q} = q_xe_x + q_y e_y$. Defect geometry is captured by the observation that rotating the vector of ${\bf q}$ at a given point by an angle $2\alpha$ rotates the `stick' of the corresponding eigendirections of $Q$ by $\alpha$. Thus if ${\bf q}$ has a singularity of winding $k$ at a point, then the eigendirections of $Q$ will have singularities of winding $k/2$---this allows us to use singularity theory to describe singularities of line fields as well as vector fields. For example, generic $\pm 1/2$ defects in a director field ${\bf n}$ correspond to Morse singularities in the associated vector field ${\bf q}$, with a local structure given by the Q-tensor,
\begin{equation} \label{eq:q_morse}
    Q^\pm_\alpha = R_\alpha \begin{bmatrix}x & \pm y \\ \pm y & -x\end{bmatrix},
\end{equation}
where the sign sets the winding $\pm 1/2$ of the singularity and $R_\alpha$ denotes the matrix encoding a rotation by an angle $\alpha$, which sets the orientation. 

The families of singularities which occur in the tomography of a director are the $A_k$ ($k \geq 1$) and $D^\pm_k$ ($k \geq 4$) families~\cite{arnold_singularity_1998}, defined in 2D by functional forms:
\begin{equation} \label{eq:singularity_classes}
  A_k : \phi = \pm x^{k+1} \pm y^2, \ \ \ \ \ \ D_k^\pm : \phi = x^2y \pm y^{k-1}.
\end{equation}
Different choices of the signs that are not related by a permutation of the coordinates correspond to different ({\it i.e.}, non-diffeomorphic) singularities. While there are many other more complex families of singularities, these simple families and their unfoldings will suffice to describe the generic structure of disclination lines, merons, and the crossing and rewiring processes of disclination lines. The $A_1$ singularities are the Morse--type singularities. An $A_k$ singularity is comprised of $k$ Morse singularities, although we emphasise that a general unfolding does not need to split the singularity into $k$ parts because it is real-valued rather than complex valued. The $D_{2k}^-$ singularities have winding $\pm 2$ and generically split into pairs of Morse singularities with the same winding. 

When discussing disclinations in line fields throughout the remainder of this text we will always consider one- or two-parameter families of 2D line fields arising from a tomography of the structure in question, and we will always describe these via the associated family of vector fields ${\bf q}$, which will need to be packaged into a Q-tensor for a realisation of the structure. In this way, the singularities $A_1$ and $D_4^-$ respectively describe disclinations, Fig.~\ref{fig3}(a), and merons, Fig.~\ref{fig3}(b), in the tomography of a director field.

\subsection{Unfoldings of Singularities Classify the Fundamental Changes of Topological Structure in a 3D Nematic}
The topology of the singular set changes when we pass over surfaces in which the projection of the director field has a singularity that changes its class. These are changes in the isotopy class, and between these surfaces the isotopy class is fixed. We identify four basic structural changes that occur in the disclinations of liquid crystals, all of which also occur in a variety of other physical systems, including optics, hydrodynamics, cosmology, geology, and biology~\cite{thom_structural_1999}. We refer to these as `structural changes' or `structural degeneracies' in the tomography. 

We now give one-parameter families of vector fields ${\bf q}$ which describe the topography in the neighbourhood of these changes. We assume the tomography surfaces are given by level sets of the $z$ coordinate, and there are $x,y$ coordinates on the surfaces. The $z$ coordinate is then the unfolding parameter of the singularity class~\cite{pollard_point_2019, arnold_singularity_1998}. When ${\bf q}$ contains a singularity of winding $\pm 2$---i.e., the projection of the director field it is describing has a winding $\pm 1$ singularity---then this must be escaped, and it and may escape either up or down. We stress that these models are the simplest models that capture the {\it topology}, and may not be the most optimal from an {\it energetic} standpoint---discussing the energetics of these structural changes is beyond the scope of this paper, and would be an interesting extension of our work. 

The most basic degeneracy is a change in the winding of a disclination that does not involve interaction with a meron. It corresponds to an $A_2$ singularity (the `fold' bifurcation~\cite{thom_structural_1999}) in the projection, Fig.~\ref{fig3}(c). 
\begin{equation} \label{eq:fold}
    {\bf q}^\text{fold} = (x^2-z)\,{\bf e}_x + y\,{\bf e}_y
\end{equation}
Geometrically, this arises exactly when a disclination line has a tangency with the tomography surfaces, as illustrated in the central panel of Fig.~\ref{fig2}. It is possible that such tangencies can be removed by a homotopy of the structure without creating, destroying, or merging any defects or merons, just as some singularities in a Morse function may be superfluous---we give an example presently. For the structure shown in Fig.~\ref{fig2} this is not the case, and this can be seen by an examination of the dividing curve on a tube around the disclination line itself, as we will discuss in Section \ref{sec:surgery} below. 

There are three additional `umbilic' degeneracies, which describe structural changes mediated by meron lines. In what follows, we take the parameter $\sqrt{z}$ to be equal to $0$ when $z \leq 0$. A meron line can split into a pair of disclinations, Fig.~\ref{fig3}(d), which is described by an unfolding of the $D_4^-$ singularity (the `elliptic umbilic' bifurcation~\cite{thom_structural_1999}):
\begin{equation}
    {\bf q}^\text{eu} = \left(\left(x+\sqrt{z}\right)\left(x-\sqrt{z}\right)-y^2\right){\bf e}_x + 2xy\,{\bf e}_y.
\end{equation}
A disclination can change its winding along part of its length by the emission of a meron line, Fig.~\ref{fig3}(e), which is described by the $D_5^\pm$ singularity (the `parabolic umbilic' bifurcation~\cite{thom_structural_1999}), where the sign determines the winding $\pm 1$ of the meron:
\begin{equation}
    {\bf q}^\text{pu} = \left(y^2-x\left(x-\sqrt{z}\right)^2\right){\bf e}_x -2y\left(x-\sqrt{z}\right){\bf e}_y.
\end{equation}
The change in the dividing curve as a result of this change is illustrated in the bottom right of Fig.~\ref{fig2}. An analogous transition may occur in the layer structure of a smectic material~\cite{machon_aspects_2019}.

Finally, there is another case where the disclination's tangent vector lies in the tomography planes, but the profile around disclinations does not change its winding, Fig.~\ref{fig3}(f). This implies that there is a meron tether attached to the disclination. This is described by an unfolding of the $D_6^+$ singularity (the `hyperbolic umbilic' bifurcation~\cite{thom_structural_1999}) and the vector field
\begin{equation}
    {\bf q}^\text{hu} = \left(y^2+x^2\left(x-\sqrt{z}\right)\left(x+\sqrt{z}\right)\right){\bf e}_x -2xy\,{\bf e}_y.
\end{equation}
There is an instability here, in which the the $D_6^+$ degeneracy can be replaced by a $D_4^-$ degeneracy as well as a pair of $A_2$ degeneracies. Geometrically this is caused by an undulation of the disclination, so that the disclination initially `bends away' from the meron (as in the elliptic umbilic case) before `bending back' via an $A_2$ fold structural change. We illustrate this later in Section~\ref{sec:self_linking}. 

The models we have given realise exactly the director fields shown in Fig.~\ref{fig3}. Non-diffeomorphic variants of each class are obtained by changing the sign of either the $x$ component or $y$ component, which will flip the winding, of changing the sign of ${\bf q}$ to rotate each director `stick' by $\pi/2$---that these give distinct classes will be important when discussing rewiring events in Section \ref{sec:rewiring}. For example, in the hyperbolic umbilic, Fig.~\ref{fig3}(f), we have illustrated $-1/2$ disclinations with a rotational meron, but we can also have $-1/2$ disclinations with a radial meron by taking $-{\bf q}^\text{hu}$, and $+1/2$ disclinations with a $-1$ meron by changing the sign of either the $x$ or $y$ component of ${\bf q}^\text{hu}$.

We have listed these degeneracies in order of complexity---the `fold' is less complex than the `elliptic umbilic', etc. Informally, the more complex the singularity the more energetically costly it is and the less likely it is to occur. This is born out by examinations of the probability distribution of degeneracies that occur during defect coarsening processes~\cite{chuang_coarsening_1993}. The fold, hyperbolic umbilic, and elliptic umbilic degeneracies all occur in observed stable textures of real materials, and we give several examples of each throughout the remainder of this work. Parabolic umbilics occur in defect coarsening and must arise as a result of defect crossings due to the necessity of a meron `tether' connecting the disclination after a crossing~\cite{poenaru_crossing_1977, alexander_entanglements_2022, kleman_disclinations_2008}. The parabolic umbilic will therefore occur as a transient part of dynamical transitions in both passive and active nematics, but we do not know of an observation of this structure as part of a stable texture in passive material. 

We also note that there is a change that can occur exclusively in the time domain, which is the conversion of a disclination line which has a constant profile $\pm 1/2$ into a disclination with constant profile $\mp 1/2$ by the emission of a meron line along the {\it entire} length, without it remaining tethered. This is as illustrated in Fig.~\ref{fig1}(d).

\section{The Surgery Perspective on Local and Global Defect Structure}
\label{sec:surgery}
In order to fully understand the relationship between isotopy, homotopy and global and local structure, we draw from another aspect of Morse theory, so-called surgery~\cite{kirby_topology_1989, gompf_4-manifolds_1999}. As we will describe in following sections, this will underpin an understanding of various structural features of disclination lines, such self-linking and charge, as well as dynamical processes such as the rewiring or crossing of defect lines, non-trivial examples of which typically involve mediation by merons. It also illustrates how homotopies localised in a neighbourhood of a disclination line may introduce structural degeneracies into the tomography. 

We describe the most basic topological features up to isotopy in terms of dividing curves. For a complete understanding, it is necessary to discuss not just point defects, merons, and disclinations, but also nonsingular textures that are either uniform or contain a cholesteric helix. Thinking of parts of the material in this way closely parallels the perspective offers by the Volterra process, and the ideas of this section offer a new perspective on results of Kleman~\cite{kleman_classification_1977, kleman_relationship_1977, kleman_defects_1989}. Once have described the isotopy classification, we exhibit the two homotopies that may change the local structure and how these impact that dividing curves: these homotopies effect the `extended Volterra process' of Kleman \& Friedel~\cite{kleman_disclinations_2008}.

\subsection{Dividing Curves for Point Defects}
In 3D, a generic point defect corresponds to a Morse singularity in the director field~\cite{pollard_point_2019}, as in Eq.~\eqref{eq:morse_singularity}. The defect can be surrounded by a sphere and the director can be given an orientation on that sphere. The dividing curve $\Gamma$ induced on that sphere will consist of some number of closed curves bounding disks, and the count of these determine the topological charge through the formula \eqref{eq:euler_class2}~\cite{copar_topological_2012, pollard_contact_2023, pollard_escape_2024}. Even if we do not choose an orientation for the director, the dividing curve still allows us to distinguish a radial hedgehog (MI 0 or 3) from a hyperbolic defect (MI 1 or 2): the former have no dividing curve while the latter have a dividing curve with two components, Fig.~\ref{fig4}(a). 

A sphere whose dividing curve contains a single component---more generally, an odd number of curves that separate the sphere into exactly two disks and an even number $2m$ of annuli---encloses a net zero defect charge, and thus we can detect the absence of charge in a region from the dividing curve without assigning a consistent orientation over the boundary of that region, a fact that we will return to in Section \ref{sec:defect_charge}.

\subsection{Dividing Curves for Merons and Twist Domains}
We describe meron lines and nonsingular twist domains on solid torii (or cylinders) by their dividing curve. A texture on the solid torus $T^2$ induces a dividing curve $\Gamma$ on the boundary which, up to isotopy, determines a homotopy class in the fundamental group $\pi_1(T^2) \cong \mathbb{Z}^2$ of the torus. By choosing a meridian curve $m$ and longitude $\ell$ on the torus any class $[\Gamma] $ can be represented as the straight line $\Gamma = a\ell + b m$ for integers $(a,b)$ that define the rational slope $b/a \in \mathbb{Q}$ of the line, and the tangent vector to the line is $b {\bf e}_\theta + a {\bf e}_\phi$ for $\theta$ the coordinate on the meridian and $\phi$ the coordinate on the longitude. The dividing curve may have multiple components, and this is codified in the homotopy class---for example, a two component dividing curve where each component is parallel to a longitude belongs to class $(2,0)$, which is induced by a uniform director field on the solid torus. In general, the number of components $|\Gamma|$ of the dividing curve in the class $(a,b)$ is the largest integer $n$ such that $a/n, b/n$ are still both integers. An empty dividing curve is also a possibility, corresponding to the class $(0,0)$. All cases may be realised by distinct isotopy classes of texture, either by a defect-free texture or else a texture containing a disclination or a meron, and in turn all {\it homotopy} classes of defect can be represented by one of these examples---there are additional isotopy classes which contain nullhomotopic components to the dividing curve. 

This is entirely analogous to the Volterra construction. The deviation of the dividing curve from the curve $(2,0)$ is a measure of distortion away from a uniform state, and different homotopy classes in  $\pi_1(T^2)$ correspond to the displacements in the Volterra process~\cite{kleman_classification_1977, kleman_defects_1989, kleman_points_1983, kleman_relationship_1977, kleman_disclinations_2008}. The constructions of this section can be generalised to materials with symmetry groups that differ from the nematic, but then then certain classes may not be realisable due not meeting the symmetry constraints (e.g., a polar material cannot have disclinations).  

\begin{figure*}
\centering
\includegraphics[width=0.98\textwidth]{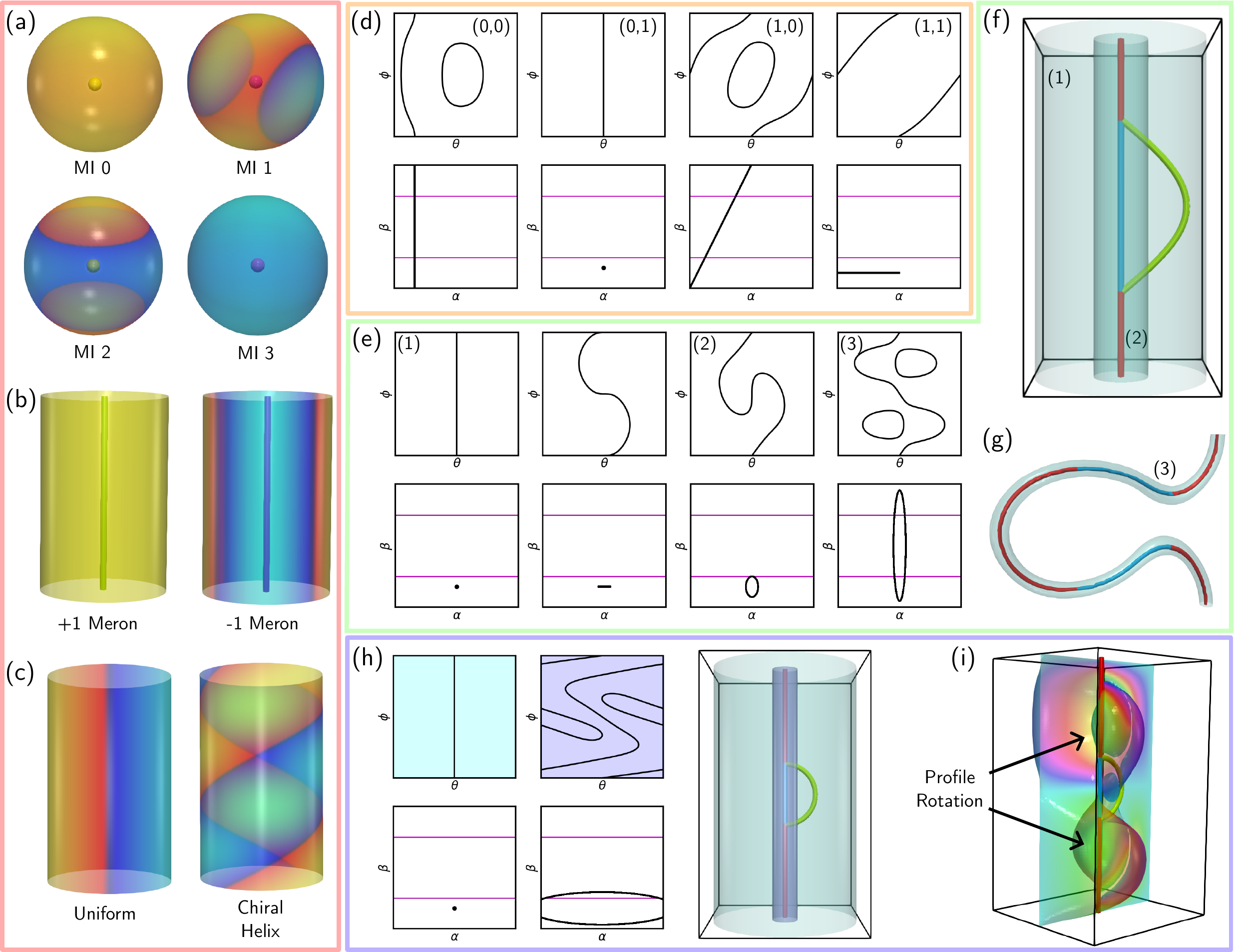}
\caption{Examining the dividing curves on a surfaces surrounding a topological feature provides a complementary perspective to tomography. Homotopies of the local structure of defects may change the isotopy class of this dividing curve, and doing so will in turn introduce structural changes into the tomography. (a-c) Orientable features are either (a) point defects (generically,of Morse type) or (b) meron lines, and we can assume a local orientation is given. The dividing curve (black) separates the surface enclosing the defect into regions $S^+$ (yellow/orange) and $S^-$ (blue/purple) where the director points respectively out of and into the surface. (c) We also consider the dividing curves on torus tubes in which the director has no defect, but may wind along the longitude of the torus as in a cholesteric---these are described by the director field ${\bf n}_{0,b}$ in \eqref{eq:nodefect}, and we illustrate $b=0$ (a uniform state) and $b=1$ (a chiral helix). (d) There are four homotopy classes of disclination lines, as described in the main text. The simplest possible dividing curves on a boundary torus associated to the class $(l, q) \in \mathbb{Z}_2 \times \mathbb{Z}_2$ are shown in the top row, for a torus parameterised by a meridian angle $\theta \in [0, 2\pi]$ and longitudinal angle $\phi \in [0, 2\pi]$. In the bottom row we we illustrate that parameters $\alpha(\phi), \beta(\phi) \in [0, 2\pi]$ that are used in the model~\eqref{eq:degennes_model} to construct director fields with these specific dividing curves, as described in the text. (e) A `protrusion growth process', a homotopy of the director field that induces a transformation of the dividing curve of a $(0,1)$ disclination. The homotopy produces two nullhomotopic components in the dividing curve. This process may occur in (f) the emission of a meron line segment, which introduces two parabolic umbilic degeneracies to the tomography, and also (g) during the growth of a protrusion in a disclination, an integral part of the emission of a disclination loop from a line segment which introduces four `fold' degeneracies. The dividing curves on the blue surfaces labelled (1), (2), and (3) in panels (f,g) are as shown in the like-labelled panels in (e). (h) The second homotopy of the local structure changes the winding of the director along a longitude, a `Hopf process'. This changes the longitudinal component of the dividing curve by introducing two additional intersections between it and the longitude, and adds an additional nullhomotopic component. The dividing curves on the blue/purple torii are those shown in the panels of the same colour. This process involves the emission of a meron tether, and hence the introduction of parabolic umbilic degeneracies. (i) A Pontryagin--Thom surface reveals the rotation of the local structure and illustrates the difference between this structure and the structure shown in panel (f), which has no rotation of the local profile. }
\label{fig4}
\end{figure*}

We consider a texture on the solid torus with a singular line at the core with nonzero winding $k \in \frac{1}{2}\mathbb{Z}$ around the meridian and winding $b \in \frac{1}{2}\mathbb{Z}$ around the longitude. For integer $k$ these lines can be escaped up or down to yield a meron, and on a cross-sectional disk this will either correspond to the element $(k,0) \in \pi_2(\mathbb{RP}^2, \mathbb{RP}^1) \cong \mathbb{Z} \times \mathbb{Z}$ if it is escape up, or to the element $(0,-k)$ if it is escape down; the dividing curve does not detect the difference. All possible singular lines can be described by
\begin{equation} \label{eq:generic_line}
    {\bf n}_{k,b} = \cos(k\theta+b\phi) \, {\bf e}_x + \sin(k\theta+b\phi)\, {\bf e}_y.
\end{equation}
where $\theta \in [0, 2\pi]$ is the polar angle and $\phi \in [0, 2\pi]$ is the coordinate along the axis of the torus. The dividing curve on the boundary torus is given by the set
\begin{equation}
    {\bf n} \cdot {\bf e}_r = \sin((k-1)\theta + b\phi) = 0.
\end{equation}
This implies that the dividing curve realises the class $(2(1-k), 2b) \in \pi_1(T^2)$. Examples for the basic merons are shown in Fig.~\ref{fig4}(b).

Defect-free textures have two intersections between the dividing curve and a meridian. All of these can be described by~\ref{eq:generic_line} with $k=0$, or equivalently by a tube cut through a cholesteric helical state:
\begin{equation} \label{eq:nodefect}
    {\bf n}_{0,b} = \cos(b\phi) \, {\bf e}_x + \sin(b\phi)\, {\bf e}_y + {\bf e}_\phi,
\end{equation}
where $b \in \frac{1}{2}\mathbb{Z}$ is half-integer, $x,y$ are coordinates in the cross-sectional disks of the torus and $\phi \in [0, 2\pi]$ is the coordinate along the axis of the solid torus. Along a longitude, the director turns by $b$ full $2\pi$ rotations, which allows us to have nonorientable textures when $b$ is fractional. They yield the classes $(2, 2b)$. Two examples are shown in Fig.~\ref{fig4}(c).

For completeness, we note these local models~\eqref{eq:generic_line} and~\eqref{eq:nodefect} can also be described in the Q-tensor picture by
\begin{equation}
    Q_{k,b} = \cos(b\phi) Q_k + \sin(b\phi) JQ_k,
\end{equation}
where $Q_k$ is 2D Q-tensor encodes a defect of winding $k$ and $J$ is the $\pi/2$ anticlockise rotation around the $\phi$ direction.

\subsection{Dividing Curves and Homotopy Classes of Disclination Lines}
Generic disclination lines are realised by setting $k=\pm 1/2$ in \eqref{eq:generic_line}. In this case the dividing curve again realises the class $(2(1-k), 2b) \in \pi_1(T^2)$. We therefore see that all homotopy classes of curve can be realised by some feature. Using the addition in this group, we can write this class as $(2,0)+(-2k,2b)$. Then $(-2k,2b)$ is a topological measure of the way that the material is distorted away from a uniform state by an isotopy. For a disclination or meron, we can define a `Burgers vector' ${\bf B}$ to be orthogonal to the dividing curve. For a straight disclination of winding $k$ with no profile winding along the longitude this results in ${\bf B} = k{\bf e}_\theta$, the same as an edge dislocation in a crystal, while a disclination with winding $b$ around the longitude will have ${\bf B} = k{\bf e}_\theta - b{\bf e}_\phi$, akin to a screw dislocation in a crystal. The Burgers vector defined by the strain---see for example Refs.~\cite{kleman_points_1983, long_geometry_2021}---is $k{\bf e}_z$ for both types, which makes {\it all} nematic disclinations akin to screw dislocations in a crystal. It would be interesting to compare the analysis of disclination dynamics in Refs.~\cite{long_geometry_2021, long_applications_2024, schimming_kinematics_2023}, which make use of the Peach--Koehler force defined using the Burgers vector, with a similar analysis based on dividing curves. 

Different choices of $k, b$ in~\eqref{eq:generic_line} yield distinct isotopy classes~\cite{pollard_contact_2023}, but not necessarily distinct homotopy classes of disclination. To understand the distinction and its consequences for the structure and dynamics of the defects, we now recall J{\"a}nich's classification of the homotopy classes of disclination lines~\cite{janich_topological_1987}. 

\begin{figure}[t]
\centering
\includegraphics[width=0.98\linewidth]{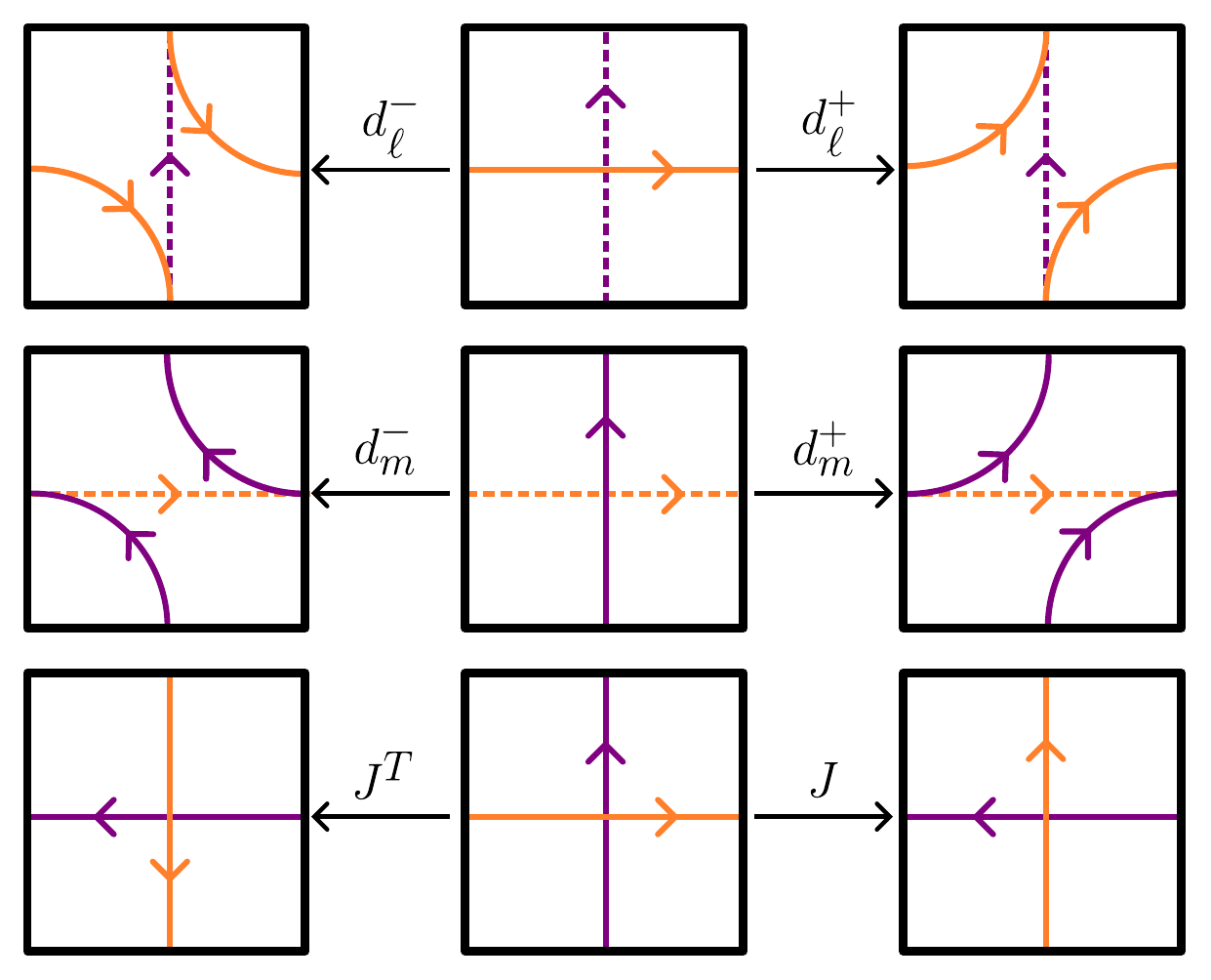}
\caption{Dehn twists on the torus allow us to enumerate the isotopy classes of disclination lines, and we will use them to illustrate how rewirings effect the dividing curve. We visualise the effect of a Dehn twist by showing its action on the meridian line (orange) and the longitude (purple) on the torus. The central column illustrates the meridian and longitudes on the initial torus, oriented so that the normal to the torus points out of the page, while the left and right columns show the result of applying the indicated transformations, defined in Eqs.~\eqref{eq:dehn_twists} and~\eqref{eq:torusJ}. Top row: Left-handed ($d^-_\ell$) and right-handed ($d^+_\ell$) Dehn twists around the longitude (dashed) leave the longitude fixed but add a twist to the meridian. Middle row: Likewise for the left-handed ($d^-_m$) and right-handed ($d^+_m$) Dehn twists around the meridian. Bottom row: The map $J$ effects a $\pi/2$ anticlockwise rotation of the entire torus, and its inverse $J^T$ effects a rotation in a clockwise sense. We keep the orientation of the meridian and longitude consistent under the rotations $J, J^T$.}
\label{fig5}
\end{figure}

Here and in the next section we make use of the theory of surface diffeomorphisms and the mapping class group~\cite{farb_primer_2012}, whose elements consist of diffomorphisms of a surface up to isotopy. The mapping class group of the torus is isomorphic to the special linear group $\text{SL}(2, \mathbb{Z})$, the set of $2 \times 2$ matrices with determinant 1. For surfaces, the Dehn-Lickorish theorem~\cite{farb_primer_2012} shows that mapping class groups are generated by a special type of map called a Dehn twist. On the torus, Dehn twists can be described by their action on the meridian and longitude as follows. Writing the tangent to the meridian as a vector ${\bf e}_\theta =(1,0)^T$ and the tangent to the longitude as a vector ${\bf e}_\phi = (0,1)^T$, the action of a Dehn twist around the meridian $d^\pm_m$ and around a longitude $d^\pm_\ell$ can be written as matrices in $\text{SL}(2, \mathbb{Z})$, 
\begin{equation} \label{eq:dehn_twists}
    \begin{aligned}
        d_m^- &= \begin{bmatrix} 1 & -1 \\ 0 & 1\end{bmatrix}, \ \ \ \ \ \ \ \ d_m^+ &= \begin{bmatrix}1 & 1 \\ 0 & 1\end{bmatrix}, \\ 
        d_\ell^- &= \begin{bmatrix}1 & 0 \\ -1 & 1\end{bmatrix}, \ \ \ \ \ \ \ \ d_\ell^+ &= \begin{bmatrix}1 & 0 \\ 1 & 1\end{bmatrix}.
    \end{aligned}
\end{equation}
The sign $\pm$ corresponds to left-handed ($-$) and right-handed ($+$) Dehn twists. There is also the $\pi/2$ anticlockwise rotation $J$, given by the matrix 
\begin{equation} \label{eq:torusJ}
    J = \begin{bmatrix} 0 & -1 \\ 1 & 0\end{bmatrix} = d^-_m d^+_\ell d^-_m.
\end{equation}
These maps are illustrated in Fig.~\ref{fig5}. To classify the local structure of a disclination up to homotopy, J{\"a}nich considered the map $f : T^2 \to \mathbb{RP}^2$ induced by the director on a torus tube surrounding the disclination. This map must be nonorientable around the meridian to describe a disclination. We choose a basic map $f_0$ with this property, and consider the family of maps 
\begin{equation} \label{eq:fj}
    f_j := f_0 \circ \left(d^+_m \right)^j.
\end{equation}
If the map $f_0$ corresponds to $(k, 0)$ in in \eqref{eq:generic_line}, then $f_j$ arises from choosing $(k, j/2)$ in \eqref{eq:generic_line}. Because these director fields are related by composing with Dehn twists, they are by definition not isotopic. However, J{\"a}nich showed that the map $f_4$ is homotopic to $f_0$~\cite{janich_topological_1987}. 

Additional perspectives on this result have been given by Chen in terms of the Pontryagin--Thom construction~\cite{chen_topological_2012}; by \v{C}opar in terms of the quaternion group~\cite{copar_topology_2014}; and by Alexander \& Kamien in terms of the Whitehead construction~\cite{alexander_entanglements_2022}. We give a proof of this result in terms of manipulations of the dividing curve on a torus surrounding the disclination, which helps to connect it to physical processes that may occur in a real material---this proof is spread over both the remainder of this section and the subsequent Section \ref{sec:self_linking}. 

J{\"a}nich's result then establishes an isomorphism between homotopy classes of directors around a disclination and the set $\mathbb{Z}_2 \times \mathbb{Z}_2$. We identify the pair $(l,q)$ classifying the line as follows: $l$ is the number of $\pi$ turns the director field makes as it moves along a longitude of the torus, modulo 2; $q+1 \text{ mod } 2$ is the evaluation of the `twisted Euler class'~\cite{machon_global_2016, bott_differential_1982} of the nonorientable director field on a torus around the disclination, which can be identified with a $\mathbb{Z}_2$ Skyrmion charge. The value $q$ (rather than $q+1$) has been identified with a defect charge~\cite{alexander_colloquium_2012, alexander_entanglements_2022}, but one must be a little more careful; we describe the distinction between the twisted Euler class and defect charge in more detail in Section \ref{sec:defect_charge} below. 

The simplest possible dividing curves achievable for the four classes $(l,q) \in \mathbb{Z}_2 \times \mathbb{Z}_2$ are shown in Fig.~\ref{fig4}(d). We produce explicit directors realising these dividing curves with a model for the local structure of a disclination introduced by Friedel \& deGennes~\cite{friedel_buckling_1969}, which has been used in many recent studies of disclination motion~\cite{binysh_three-dimensional_2020, long_geometry_2021, schimming_kinematics_2023}. We do this in order to help connect our work to these previous studies, although emphasise that the description here is a topological result and does not depend on a particular realisation. Denote by ${\bf t}$ the tangent vector to $K$, and define Cartesian coordinates $u, v$ on the disks orthogonal to $K$ in some small toroidal neighbourhood. Associated to these Cartesian coordinates we have polar coordinates $r, \theta$ on the orthogonal disks and a a coordinate $\phi$ along $K$. We choose a `rotation vector' $\Omega$ along $K$ with the property that the director lies everywhere in the plane orthogonal to $\Omega$, and then define ${\bf n} = \cos(\theta/2){\bf m} + \sin(\theta/2){\bf m}_\perp$, where
\begin{equation} \label{eq:degennes_model}
  \begin{aligned}
    {\bf m} &= \cos \alpha \ {\bf e}_u + \sin \alpha \ {\bf e}_v, \\
    {\bf \Omega} &= \cos \beta \ {\bf t} - \sin \beta \ {\bf t} \times {\bf m}, \\
    {\bf m}_\perp &= \cos \beta \ {\bf t} \times {\bf m} + \sin \beta \ {\bf t},
  \end{aligned}
\end{equation}
The angles $\alpha, \beta$ are functions that parameterise the model and control the shape of the director. A profile (the projection of the director onto a disk of constant $\phi$) is called `splay' if $\beta = 0$ or $\pi$, and `twist' if $\beta = \pi/2$ or $3\pi/2$. A twist profile is `radial twist' if $\alpha = 0$ or $\pi$. and `tangential twist' if $\alpha = \pi/2$ or $3\pi/2$. In general $\alpha, \beta$ are functions of the coordinates $r,\theta,\phi$ in a neighbourhood of the line. We are considering an arbitrarily small neighbourhood of the line, and hence make a simplification where they are functions of $\phi$ alone. Thus, the pair $\alpha(\phi), \beta(\phi)$ gives a map $S^1 \to T^2$; we visualise the image of these maps in the lower row of Fig.~\ref{fig4}(d). This illustrates the connection between the structure of the dividing curve and the more common characterisation of the profile in terms of `splay' and `twist' regions.

In each case there exists a component of the dividing curve which has a nonzero intersection with any meridian on the torus: we refer to this as the `longitudinal component'. There may be additional components which bound disks on the torus: these are `nullhomotopic components'---they can be shrunk down to a point on the torus, as opposed to curves that wrap around a meridian or longitude and cannot be shrunk to a point---and they are associated with the presence of regions of `twist' profile. The number of times the dividing curve intersects the boundary of a meridonal disk $D$ determines the winding of the projection of the director into $D$. One intersection corresponds to a $+1/2$ profile and three intersections to a $-1/2$ profile, and the absence of nullhomotopic components implies the profile is constant, independent of the disk $D$ we use to measure it~\cite{pollard_contact_2023}.

\subsection{There are Two Fundamental Homotopies That Change the Isotopy Class of The Dividing Curve}
We now describe how homotopies of the local structure of a disclination line change the dividing curve, and how these relate to global topological and geometric changes. There are two fundamental homotopies, from which more complex changes in structure are obtained by performing these basic homotopies multiple times. 

The first kind of homotopy does not change the homotopy class of the longitudinal component of the dividing curve in $\pi_1(T^2)$, but either adds or subtracts a pair of nullhomotopic loops, Fig.~\ref{fig4}(e). This process can occur in one of two ways. Firstly, it is possible for the disclination to emit a meron line along part of its length, Fig.~\ref{fig4}(f). Alternatively, the disclination can undergo a change in its geometry by developing a protrusion, Fig.~\ref{fig4}(g). This second case is part of the process by which a disclination line emits a loop, which we discuss in more detail in Section \ref{sec:rewiring} below. We refer to this homotopy as a `protrusion growth process', because it is more naturally associated with the geometric change illustrated in Fig.~\ref{fig4}(g) rather than the emission of a meron tether.

The second kind of homotopy applies a pair of Dehn twists, either left $d^-_m$ or right $d^+_m$, to the longitudinal part of the dividing curve, and either adds or subtracts a single nullhomotopic loop. This change is illustrated in Fig.~\ref{fig4}(h). This cannot arise from a change in the geometry of the line, it must result from the emission of a meron tether, or an emission of an entire meron loop from the line---a similar change can occur by rewiring of disclination lines, see Section \ref{sec:rewiring} below, but this is not a homotopy of the local structure. Locally, changing the self-linking results from a rotation of the profile and, for a $+1/2$ line, must occur in a series of Hopf bifurcations. The global structure of the Pontryagin--Thom surface~\cite{chen_topological_2012} where the $x$ component of the director field vanishes is shown in Fig.~\ref{fig4}(i), coloured according to the angle between the other two components of the director. We refer to this as a `Hopf process', because it involves Hopf bifurcations in the disclination profile.

\begin{figure*}[t]
\centering
\includegraphics[width=0.98\textwidth]{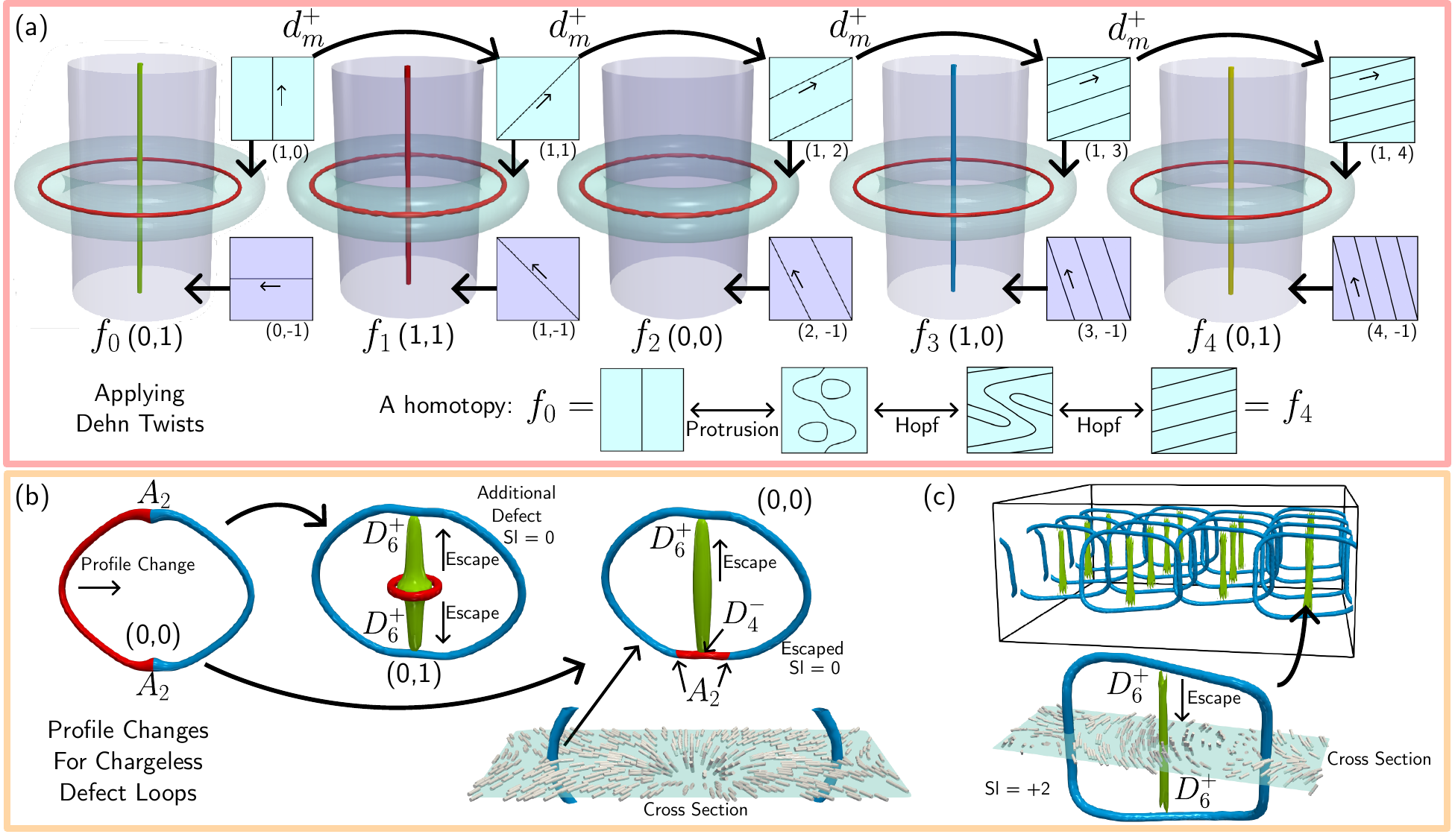}
\caption{Changes in the self-linking number of a disclination line result from interactions between the disclination and either other disclination lines or merons. This is captured by applying Dehn twists to the torus tube that encloses a given disclination, which can be seen to correspond to removing $+1/2$ disclination lines from the tube's complement. (a) Beginning from a $(0,1)$ defect loop with everywhere $+1/2$ profile (leftmost panel), we consider the distinct textures that result from applying Dehn twists to the dividing curve (moving rightwards). Our loop lives inside a solid torus tube (pale blue), and initially induces a map $f_0 : T^2 \to \mathbb{RP}^2$ on the boundary of this tube. The complement of this tube in $S^3$ is another solid torus (purple) whose boundary is glued to that of the pale blue torus via the map $J$ that associates a meridian on the pale blue torus to a longitude on the purple torus. This is the Heegaard decomposition~\cite{heegaard_preliminary_1898}. By fixing a dividing curve on the blue torus, we therefore determine a dividing curve on the purple torus---in each panel we show the minimal structure within the purple torus which achieves the required dividing curve. Each time we applies a Dehn twist $D_m^+$ to the dividing curve we obtain a texture corresponding to a map $f_j$ as in Eq.~\ref{eq:fj}. This changes the J\"anich invariant from $(0,1)$ to $(1,1)$ for $f_1$, $(0,0)$ for $f_2$, $(1,0)$ for $f_3$, and finally $f_4$ corresponds to the homotopy class $(0,1)$ again, but now in a distinct isotopy class from $f_0$~\cite{janich_topological_1987}. The homotopy that connects them is shown in the lower part of the panel, and involves the two processes illustrated in Fig.~\ref{fig4}. In each case the dividing curves on the boundaries of the two torii are illustrated along with the homotopy class in $\pi_1(T^2)$ they represent. In the complement (purple) we show $+1$-winding merons in green and $-1$-winding merons in yellow. (b) A basic loop from J\"anich class $(0,0)$ with no self-linking has half of its profile $+1/2$ winding (red) and half $-1/2$ winding (blue), and is described by an unfolding of $A_2$. We can convert the $+1/2$ winding to $-1/2$ by pulling out a $+1$ radial meron line (green). Depending on whether this escapes up everywhere, or escapes up in one half and down in the other, this may result in either another $(0,0)$ loop or a $(0,1)$ loop with an additional point defect. The singularity types classifying the structural degeneracies on the loop are indicated, and self-linking of all of these lines vanishes. (c) By rotating the profile we create a geometrically distinct kind of loop described a different form of the $A_2$ singularity. Performing the same process here results in a $+1$-rotational meron. The direction of escape sets the handedness~\cite{pollard_escape_2024}; we illustrate escape down, for a right-handed texture. The loop then has $-1/2$-profile everywhere and has self-linking $+2$. Lattices of these loops have been observed in simulations of cholesterics in a thin domain~\cite{fukuda_ring_2011, fukuda_quasi-two-dimensional_2011, kwok_cholesteric_2013, nych_spontaneous_2017}, and the cross-section is exactly that of a meron lattice. }
\label{fig6}
\end{figure*}

\section{The Dividing Curve Determines the Orientation and Self-Linking of Disclination Lines}
\label{sec:self_linking}
Now that we understand the two fundamental types of local homotopies that change the isotopy class, we can discuss the geometric and dynamical aspects of these transitions and the relationship with J{\"a}nich's theorem. This involves changes in the self-linking number of a disclination, which we describe now in terms of Dehn twists applied to dividing curves. This section has much overlap with the work of \v{C}opar \& \v{Z}umer on nematic braids~\cite{copar_nematic_2011, copar_PRE_2011, copar_quaternions_2013}, where the topological self-linking of disclinations was originally defined, and the observed relationship between self-linking in colloidal systems and the actual geometry of the disclination lines that appear in these systems is described in some detail. See also the review in Ref.~\cite{copar_topology_2014}. There are still many open questions about the relationship between topology and geometry in nematics, which we do not discuss here but represent an important extension of this work. 

To properly describe linking and self-linking of disclination lines it is first necessary to orient them in their tangent direction. While merons are naturally oriented by the director field, disclination lines are not. Instead, the set of disclination lines inherits a consistent orientation from the manifold $M$ on which the director is defined. The manifold has an orientation defined by a choice of a (local) ordered frame field for its tangent space---generally we take it to be right-handed. Any submanifold can then be then be assigned an orientation, coming from a choice of frame along the surface that is also right-handed. Given a solid torus tube enclosing a disclination, this torus tube is then oriented as a submanifold of $M$, and its central line (the disclination) in turn receives an orientation. Alternatively, we may choose a Seifurt surface for the set of disclinations and orient them as the boundary of the Seifurt surface; this either gives the same orientation or the opposite orientation to that given by the manifold. The tube around a meron can be oriented by the manifold in this way, and then it possible to consistently assign merons elements of the homotopy group $\pi_2(\mathbb{RP}^2, \mathbb{RP}^1)$~\cite{machon_global_2016, pollard_escape_2024}, which realises them as `escape up' or `escape down' relative to the manifold-induced orientation of the line, even when the director itself is not oriented.

The dividing curve $\Gamma$ on a boundary torus then endows each disclination $K$ with a framing in the following way. Take a component $\gamma_\ell$ of $\Gamma$ that is longitudinal and not nullhomotopic---if there are multiple such components, then they all belong to the same homotopy class in $\pi_1(T^2)$ and the choice is irrelevant. Since $\gamma_\ell$ is an element of $\pi_1(T^2)$ it can be viewed as an oriented curve in the complement of $K$, which means we have a well-defined self-linking number $\text{SL}(K) := \text{Lk}(K, \gamma_\ell)$. We then endow $K$ with a framing corresponding to this self-linking number, which can be seen simply a choice of normal vector which points at $\gamma_\ell$ everywhere. This is a purely topological way of making $K$ into a ribbon, and is entirely consistent with the geometric methods introduced by \v{C}opar \& \v{Z}umer\cite{copar_nematic_2011, copar_PRE_2011, copar_quaternions_2013, copar_topology_2014}; their definition involves a constant multiplicative factor which is superfluous from a topological point of view, but otherwise our definitions agree. Other tensors derived from the director gradients can be used to define a dividing curve and the self-linking: for example, the eigendirections of the orientation tensors defined by Long \textit{et al.}~\cite{long_geometry_2021} naturally trace out a set of curves on the torus surrounding a disclination which are equivalent to the dividing curve, as will any other reasonable tensorial measures of winding that can be extracted from $\nabla {\bf n}$. 

The self-linking of a disclination is not a homotopy invariant of a disclination line because it can be changed via the Hopf process illustrated in Fig.~\ref{fig4}. However, its mod 2 reduction is a homotopy invariant, because an even (respectively odd) self-linking implies that the disclination line links with an even (respectively odd) number of other disclinations, where merons are counted as a pairs of disclinations~\cite{janich_topological_1987, alexander_colloquium_2012, copar_topology_2014}. Rewirings may or may not change the self-linking number, which we discuss presently in Section \ref{sec:rewiring}. Crossing a disclination line through another disclination changes both of the self-linking numbers by $\pm 1$ and changes the homotopy type accordingly, as we describe in Section \ref{sec:crossing} below. 

We can use these observations about the self-linking to give a proof of J\"anich's classification~\cite{janich_topological_1987} based on manipulations of the dividing curve, which helps to visualise this difficult topological result and understand its physical meaning. In Fig.~\ref{fig6}(a) we show the result of repeatedly changing the self-linking of a disclination $K$ via applied Dehn twists. We visualise this process using the Heegaard decomposition of $S^3$~\cite{heegaard_preliminary_1898, kirby_topology_1989}. This realises the 3-sphere as the union of two solid torii $H_1, H_2$, which are glued together along their common boundary. The map that effects the gluing is the map $J$ defined in \eqref{eq:torusJ} which attaches the longitude on one torus to the meridian on the other. We regard our disclination $K$ (red) as being surrounded by $H_1$, shown in pale blue in Fig.~\ref{fig6}(a). Whatever structure exists in the complement can be reduced to a single curve, by `squeezing' all of the defects down onto the core line of the solid torus $H_2$, which we show in purple. The dividing curves on the boundaries of these solid torii must be related by the map $J$ in order for the textures to `glue' correctly. The defect shown at the centre of $H_2$ is therefore the minimal defect structure necessary to complete the texture in $H_1$ to a smooth texture over $S^3$. 

We begin with a case where $K$ belongs to the J\"anich class $(0,1)$, with constant $+1/2$ profile, which corresponds to a map $f_0 : T^2 \to \mathbb{RP}^2$. Its complement is a meron line of winding $+1$ (green). We show the dividing curves as insets, labelled by the class in $\pi_1(T^2)$ they correspond to; these curves are also oriented (black arrows), so that $(1,0)$ is different from $(-1,0)$. Applying the Dehn twist $d^+_m$ to the dividing curve of $K$ changes its J\"anich class to $(1,1)$, corresponding to a map $f_1 = f_0 \circ d_m^+$. This process is topologically equivalent to excising a $+1/2$ defect from the complement via surgery, which then converts a $+1$ meron into a $+1/2$ disclination. In this example $K$ (in the blue tube) is oriented anticlockwise and has self-linking $+1$; the other disclination (in the purple tube) is oriented `up' the page and has self-linking $-1$. 

\begin{figure*}[t]
\centering
\includegraphics[width=0.98\textwidth]{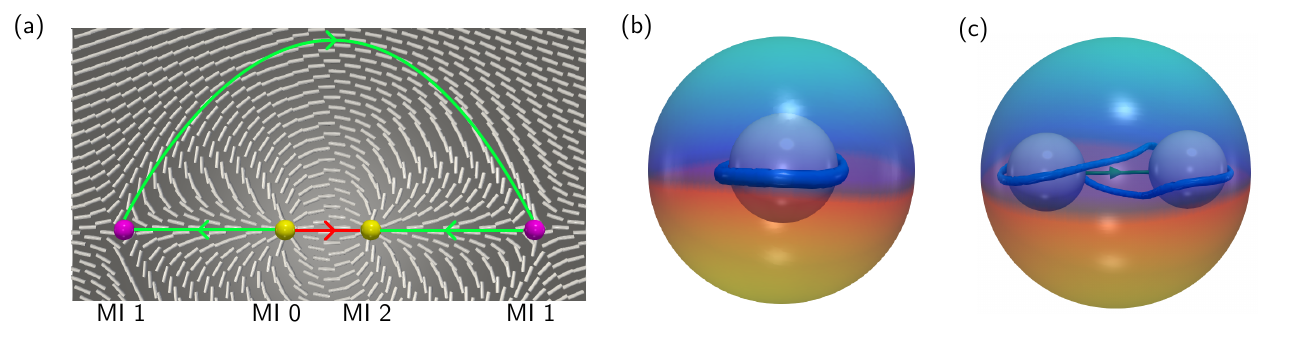}
\caption{Care must be taken when assigning charge and Morse Indices to defects in order to ensure local as well as global consistency, which is essential for correctly describing the results of interactions between defects. (a) In a 2D orientable line field, the choice of a Morse Index (MI) for any of the singular points assigns a Morse Index to every other singular point by propagating the choice along integral curves. Equivalently, we may orient a single integral curve, and propagate this orientation along the other integral curves. Either case works and assigns a unique (up to global sign flip) charge even in the presence of nonorientable singularities. (b) The dividing curve on a sphere surrounding a Saturn ring---a disclination encircling a colloidal particle with normal anchoring of the director field on the boundary---illustrates that this pair has vanishing defect charge. Thus, if we decide the colloid represents a Hedgehog with MI 0 and charge $+1$, then the disclination has charge $-1$ and must contract to a hyperbolic point defect of MI 1. We have oriented the director over the sphere---yellow is outwards, blue inwards---but the fact that it contains zero defect charge is independent of the orientation as is seen from the dividing curve (black). (c) For more complex systems we may orient point defects, or particles that represent point defects, as described in panel (a) by orienting integral curves that connect them (green). The two colloids illustrated must therefore have opposite charges. Since the dividing curve shows the entire structure has vanishing charge, this implies the disclination that encircles the colloids also has vanishing charge.}
\label{fig7}
\end{figure*}

Performing this process again applies a second $d^+_m$ to the dividing curve, resulting in the map $f_2$. $K$ now becomes a disclination with J\"anich class to $(0,0)$ and $\text{SL}(K) = +2$. This excises a second $+1/2$ disclination from the complement, which is then readily seen to be a chiral helix described by the director \eqref{eq:nodefect} with winding $b=-1/2$. In particular, we note that this disclination has no defect charge, because its complement is defect-free---we return to this observation and explain it in greater detail in Section~\ref{sec:defect_charge}. Applying a third Dehn twist to obtain the map $f_3$ is equivalent to removing another $+1/2$ disclination from the defect-free purple torus, so it leaves behind a $-1/2$ disclination with self-linking $-1$. $K$ now belongs to J\"anich class to $(1,0)$ and has $\text{SL}(K) = +3$. 

Applying yet another Dehn twist results in $K$ having a dividing curve $(1,4) \in \pi_1(T^2)$, which corresponds to a map $f_4 = f_0 \circ {d^+_m}^4$; the complement now contains a $-1$-winding meron line (yellow). J\"anich's classification asserts that this is homotopic to $f_0$, so this line must be of class $(0,1)$---it is a distinct isotopy class within the same homotopy class. We can see this clearly by using the visualisation of the allowed homotopies in Fig.~\ref{fig4}, and their effect on the dividing curve. The dividing curve has self-linking number $+4$. We can therefore apply two of the Hopf process homotopies, Fig.~\ref{fig4}(h), to obtain a new dividing curve with vanishing self-linking but two additional nullhomotopic components, exactly as shown in Fig.~\ref{fig4}(e), panel (3). Applying the protrusion growth process then carries us back to a dividing curve with a single component parallel to a longitude, exactly the original map $f_0$. This homotopy is shown schematically in the bottom part of Fig.~\ref{fig6}(a). A similar process allow us to reduce each of the structures shown in Fig.~\ref{fig6}(a) to their simplest representatives shown in Fig.~\ref{fig4}(d), which belong to different isotopy classes within the same homotopy class. 

The Hopf process is only a global homotopy if the singular line that is pulled out escapes one way---if it escapes in opposite directions then this process is instead equivalent to pulling a point defect out of the disclination, changing its J\"anich class. In Fig.~\ref{fig6}(b, c) we visualise this for a disclination belong to J\"anich class $(0,0)$. This disclination carries no charge and the director may be made fully planar, described exactly by a pair of `fold' structural degeneracies put together. We consider the director defined by 
\begin{equation} \label{eq:a2_loop}
    {\bf q}_{A_2} = \left(x^2-\cos z \right){\bf e}_x + y\, {\bf e}_y,
\end{equation}
with $z \in [-\pi, \pi]$. This is shown in Fig.~\ref{fig6}(b). Now we pull a $+1$ singular line out of the $+1/2$ region to change its winding to $-1/2$. Two outcomes result. In the first case, we escape this singular line alternately up and down, which produces a point defect in the centre; in simulations this generically grows into a small ring, as shown in Fig.~\ref{fig6}(b). The J\"anich class of the original loop has now changed to $(0,1)$. Alternatively, we can escape the singular line everywhere in the same direction. The J\"anich class remains the same, and the loop still has a small $+1/2$ region. Now the $D_6^+$ degeneracy at the bottom of the loop is instead replaced by a $D_4^-$ degeneracy along with two $A_2$ degeneracies, a possibility we described in Section~\ref{sec:tomography}. This ultimately drives a buckling of the disclination loop, changing its shape from a circle to a cardioid. 

In neither case does the self-linking change. In fact, whether the self-linking can change here is dependent on the orientation of the profile of the $(0,0)$ loop, an interesting geometric constraint. Suppose instead that that the local profile of our loop has director defined by $-{\bf q}_{A_2}$, for ${\bf q}_{A_2}$ as in Eq.~\eqref{eq:a2_loop}. Pulling out a $+1$ singular line from the $+1/2$ segment and escaping now results in a rotational meron, which can escape to give an everywhere $-1/2$ winding for the singular line, Fig.~\ref{fig6}(c). Examining the dividing curve on a torus tube shows it has self-linking $\pm 2$, as in the disclination corresponding to the map $f_2$ in Fig.~\ref{fig6}(a), with a sign according to the direction of escape. Rotational profile $+1$ singular lines have an innate chirality, whereby `escape up' corresponds to a left-handed texture and `escape down' to a right-handed texture~\cite{pollard_escape_2024}. Here, this chirality manifests itself in the sign of the self-linking number, equivalently, whether this process results in a left- or right-handed Dehn twist being applied to the dividing curve---we illustrate a right-handed texture, escape down, with self-linking $\text{Sl} = +2$.   

Disclinations with nonzero self-linking number form entangled around colloidal particles~\cite{tkalec_reconfigurable_2011, copar_topology_2014, musevic_liquid_2017}. They also arise naturally in confined cholesterics in thin cells where the pitch length and the height of the cell are incommensurate~\cite{fukuda_ring_2011, fukuda_quasi-two-dimensional_2011, kwok_cholesteric_2013, nych_spontaneous_2017}. An example of the second case, a lattice of disclination loops in a right-handed cholesteric cell\cite{fukuda_ring_2011}, is shown in Fig.~\ref{fig6}(c). This is obtained by taking the cholesteric pitch length $p = 0.75 h$, where $h$ is the height of the cell along the $z$ direction. Each disclination has everywhere $-1/2$ profile and exhibits a $+1$-meron tether with rotational profile (a double-twist cylinder), which is exactly the structure we have just described. One of these disclinations consists of a pair of $D_6^+$ `hyperbolic umbilic' structural degeneracies, exactly as shown in Fig.~\ref{fig3}(f). The merons are necessary in order for the projection into surfaces of constant $z$ to meet the constraints of the Poincar\'e--Hopf theorem. An alternative texture, not shown here but described in previous studies~\cite{kwok_cholesteric_2013, nych_spontaneous_2017}, has helices of $-1/2$ disclinations running orthogonal to the $z$ direction instead of rings---these helices also have nonzero self-linking and have $+1$ merons tethered to them, and can be seen as the result of rewiring parts of the loops in the texture in Fig.~\ref{fig6}(c).

\section{Assigning Defect Charge to Disclinations}
\label{sec:defect_charge}
In this section we discuss the question of how to assign charge to disclination loops. An assignment of charge to defects (both points and lines) must not only meet the global topological constraint imposed by the Poincar\'e--Hopf theorem, but also a local consistency constraint which governs the interactions between defects. The charge of a single defect is determined from a sphere surrounding just that defect; local consistency requires examining a sphere that surrounds a pair of defects and ensuring that the signs of their individual charges are consistent with the charge we measure across the whole sphere. Using Morse theory, it is straightforward to assign charges to a system with only point defects so that those charges are both locally and globally consistent. However, up until now, how to properly extend this to defect loops has remained unclear. In particular, a lack of appreciation for the local consistency requirement has lead to some errors in the literature. For example, whilst the conflation of Skyrmion charge with defect charge is reasonable in an oriented system [{\it viz.} Eq.~\eqref{eq:euler_class2}], in the presence of disclination lines these concepts need to be carefully disambiguated. In the following, we use surgery theory and the notion of dividing curves to set out a consistent notion of assigning charge to (arbitrary) disclination loops, correcting errors in the literature in the process.

\subsection{Morse Theory Assigns Charge to Point Defects in a Manner That is Consistent Both Locally and Globally}
First we briefly recall the definition of the charge of a point defect and its relationship with the dividing curve. Suppose we have an oriented (or at least orientable) director field in some domain. It may have only point defects. Let $W$ be the complement of small balls around each defect, regions which we interpret as the defect cores. The Euler class $e({\bf n})$ of the director field is a homotopy invariant associated to the oriented vector bundle consisting of the planes orthogonal to the director on $W$~\cite{bott_differential_1982, geiges_introduction_2008, copar_topological_2012, machon_umbilic_2016, pollard_contact_2023, pollard_escape_2024}. Its value on a surface $S$ depends only on the homotopy class $[S] \in \pi_2(W)$. The value of the Euler class on a sphere is twice the defect charge within the sphere. As Eq.~\eqref{eq:euler_class2} shows, its value on a surface that is not a sphere is twice the Skyrmion charge across that surface, that is, a signed count of the number of merons the surface intersects. The Poincar\'e Hopf theorem imposes the global constraint that the defect charges must sum to zero (or to an appropriate value if $M$ has a boundary, e.g. $+1$ for a sphere with outwards-facing radial anchoring).

A sphere $S$ that is not nullhomotopic in $W$ must surround some number of point defects, and the value $e({\bf n})[S]$ of the Euler class on the homotopy class $[S]$ gives the defect charge contained within $S$. In particular, by using the formula~\eqref{eq:euler_class2} we may compute the defect charge using the dividing curve. But the dividing curve tells us more than just the defect charge: it also tells us whether the Morse Index of the defect is 0 or 3, or whether it is 1 or 2; in the former case the sphere has empty dividing curve, in the latter it has two components, Fig.~\ref{fig4}(a). 

If the director field is not oriented then the charge is only determined up to a sign, and we also cannot tell the difference between MI 0 and MI 3. However, the requirement for both charge and MI to be locally consistent assures that are only two ways of assigning them across the material. Suppose a region of the material contains two generic point defects. We know each defect has charge $\pm 1$, but we do not known from just the local structure of a given defect which sign to assign it. However, suppose we now surround these two point defects with a sphere $S$, and that the charge across this sphere is zero. We conclude that the two point defects within $S$ must have charges of opposite sign. This is what we mean by local consistency of defect charge, and it can be extended to all defects in the system by examining all pairwise interactions in this way. Once the sign of the charge for one defect is chosen, all the rest are fixed by the requirement of local consistency. 

We can use Morse theory to simplify the computations. We illustrate this in a 2D setting in Fig.~\ref{fig7}(a), but it is equivalent in 3D. We are free to assign a MI to any point singularity. This is equivalent to choosing an orientation for one integral curve emanating from that singularity (red). This choice then propagates along integral curves of the director field (light green) and consistently assigns a MI to each other singularity---this imposes a local consistency of Morse indices, which in 3D will lead to a local consistency of defect charges. We see that the only choice is the orientation of a single integral curve, and there are therefore exactly two distinct possible assignments of Morse indices, which differ mod 2. In a 3D system this approach still works to consistently assign a Morse index---and hence charges---to the point singularities, even if there are disclination lines in the system as well.

\subsection{The Charge of a Disclination in a Colloidal System is Determined by Dividing Curves}
Now suppose the director contains some disclination loops as well as point defects. We may still define $W$ to be the complement of the defect set, and the director defines a plane field everywhere on $W$. The definition of the Euler class requires the vector bundle to be orientable, but if there are disclination lines this is not the case. Instead, we may define a `twisted Euler class'~\cite{machon_global_2016, bott_differential_1982}, which for a single disclination lives 
in the $\mathbb{Z}_2$ cohomology group $H^2(W, \mathbb{Z}_2)$, and hence its value on any surface is either $0$ or $1$, a $\mathbb{Z}_2$ Skyrmion charge. This has been identified with a defect charge~\cite{alexander_colloquium_2012, alexander_entanglements_2022}, but we should be careful: the charge of disclination line must be understood as an integer assigned to a sphere around the disclination, not to a torus, and while the Skyrmion charge in a nonorientable system is only a mod 2 integer in $\mathbb{Z}_2$, the defect charge is still an integer in $\mathbb{Z}$.

Dividing curves again provide a framework for understanding this. We first explain this in terms of a well-studied experimental system in which disclinations are nucleated around colloids~\cite{tkalec_vortexlike_2009, copar_nematic_2011, copar_PRE_2011, tkalec_reconfigurable_2011, musevic_liquid_2017}, before turning to the general case. In this colloidal system the process of assigning charge has caused some confusion: for example, in Ref.~\cite{tkalec_vortexlike_2009} a chargeless $-1$-winding meron line is incorrectly identified as carrying defect charge $-2$.

Consider the Saturn ring configuration shown in Fig.~\ref{fig7}(b). The structure consists of a colloidal particle with radial anchoring (gray), which we identify with a hedgehog point defect, as well as a disclination loop with $-1/2$ winding everywhere (blue). Surrounding the pair with a surface and computing the dividing curve, we find that together the pair must have zero charge. Therefore, assigning the colloidal particle charge $+1$ (MI 0) unambiguously assigns the defect loop charge $-1$, and vice-versa. The two defects must be able to annihilate with one another to produce a defect-free texture---if we shrink the loop down to a point defect then it must be hyperbolic with MI 1, a texture which is also seen in experiments~\cite{poulin_novel_1997, joanny_nematic_1997, musevic_liquid_2017}.  

If we consider two such colloid/disclination pairs, then the requirement that the Morse indices be locally consistent with one another makes it clear that one colloid must be MI 0 with charge $+1$ while the other must have MI 3 and charge $-1$---taken together, the pair of colloids have charge zero, and we could also see this by examining the dividing curve on a sphere that surrounds just the two colloids and not the disclinations. Additionally, each individual colloid/disclination pair is identical to the structure shown in Fig.~\ref{fig7}(b), and so the pair also has charge zero; it follows that the disclinations must have opposite charges to one another. Indeed, if these disclinations were collapsed to points they would be hyperbolic point defects with MIs 1 and 2, and a cross-section of the structure would be topologically equivalent to that shown in Fig.~\ref{fig7}(a). 

For a further example with two colloids, we see that the `figure eight' loop in Fig.~\ref{fig7}(c) has vanishing defect charge, unambiguously---this structure has also been incorrectly assigned a charge of $-2$. To see this, note that Morse theory implies the two colloids must have opposite charge, as can be seen by propagating the assignment of MI 0 on the left colloid to the right colloid along the green integral curve, and since the entire structure has charge zero, so must the disclination. Indeed, if we begin from a pair of Saturn ring configurations, the figure eight loop in Fig.~\ref{fig7}(c) is obtained by a rewiring process that merges sections of the two disclinations. Thus, we see that the charge we assign here is properly additive, exactly as it is for point defects and as is required by topology. The $-1$ meron described in Ref.~\cite{tkalec_vortexlike_2009} is obtained by merging the two Saturn ring disclinations along their entire length, and thus it is clear that it must also have zero charge. 

This reasoning extends to general configurations with $n$ colloidal particles. Any disclination entangled around some number $n$ of colloids can be built from a basic texture consisting of $n$ Saturn rings by successive rewiring processes~\cite{tkalec_reconfigurable_2011, copar_nematic_2011, copar_PRE_2011}. To determine the charge of any disclination in this kind of system, we need only make use of the following three observations: (1) a Saturn ring always has opposite charge and complementary MI to the colloid it encircles, (2) the colloids must have alternating MIs 0 and 3 and corresponding alternating charges, and (3) the charge is additive under rewiring process. 

\begin{figure*}[t]
\centering
\includegraphics[width=0.98\textwidth]{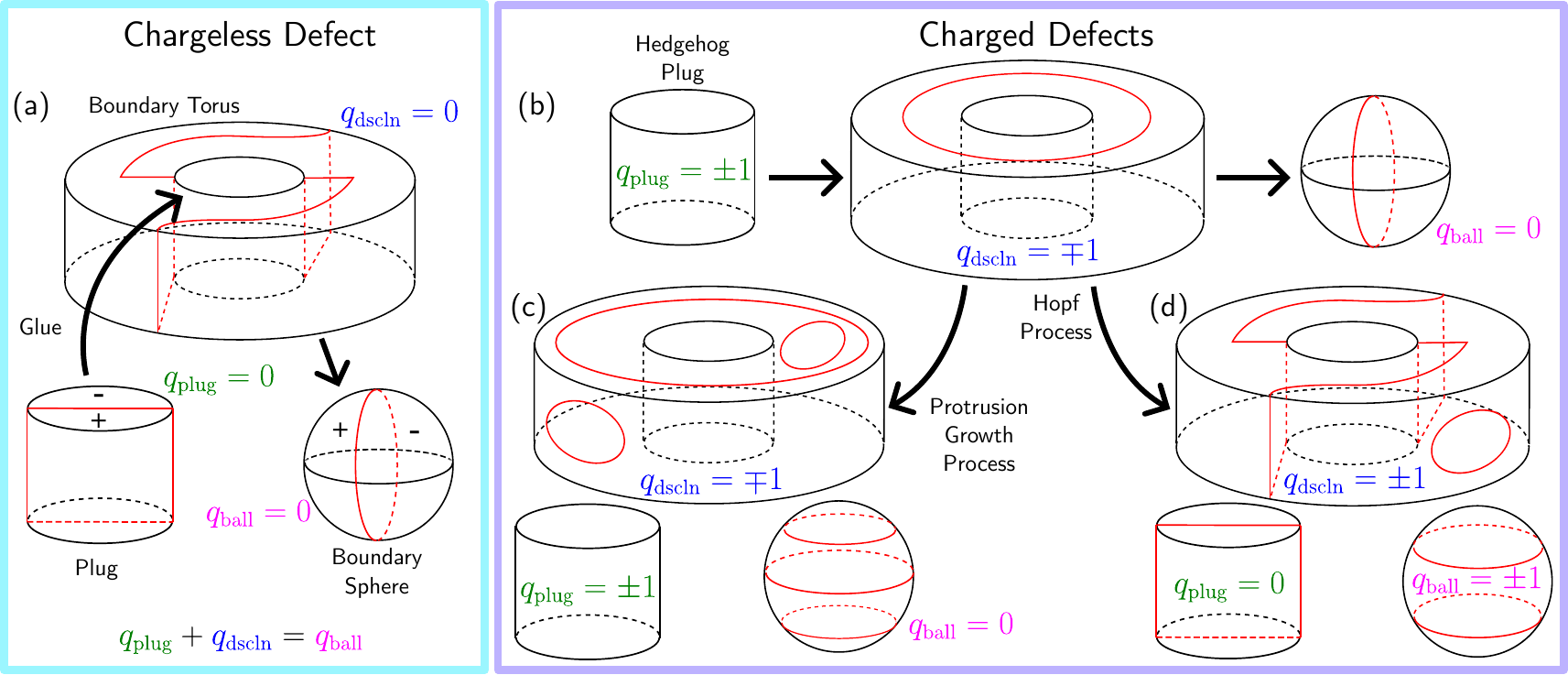}
\caption{The charge of a general disclination loop can be determined via surgery, a process equivalent to shrinking the loop down to a point defect. The associated point defect can then be assigned a Morse index consistent with the local and global constraints imposed by the other defects. (a) Consider a disclination belonging to J\"anich class $(0,0)$ with self-linking $-2$. A solid torus tube $H$ with boundary torus $T$ around this line is shown in black, with the dividing curve in red. Topologically, we can view the solid torus tube as a cylinder with a smaller cylinder cut out. To fill this hole, we can glue in a small cylinder $C$ which we call the `plug'. The texture on this cylinder must be chosen so as to attached to the dividing curve on $T$---in this case, it has a single component. After gluing $C$ into $T$ we obtain a (topological) solid ball with spherical boundary $S$ containing the original disclination as well as whatever defects were in the plug. Thus, the defect charge $q_\text{ball}$ within $S$ satisfies $q_\text{plug}+q_\text{dscln}=q_\text{ball}$, where $q_\text{plug}$ is the defect charge within the plug and $q_\text{dscln}$ is the charge carried by the disclination; we therefore determine the charge of the disclination up to a sign choice, in this case $q_\text{dscln}=0$. We can in principal orient the director over the plug $C$ (indicated by $\pm$ signs here) and impose a consistent orientation on $S$ to fix the sign. (b) The same process for a disclination from J\"anich class $(0,1)$ with self-linking $0$ show that it carries a charge $q_\text{dscln}=\pm 1$. We also illustrate the change in the dividing curve under the two homotopies, (c) the protrusion growth process and (d) the Hopf process. The charge is still $q_\text{dscln}=\pm 1$. The sign can only be fixed by global considerations, as described in the text.}
\label{fig8}
\end{figure*}

\subsection{Surgery Theory Allows for Consistent Charge Assignment for any Defect Loop}
Now we generalise this idea to arbitrary sets of defect loops in arbitrary systems, with or without point defects. The disclinations can have any knot type as long as they are unlinked with other disclinations. Consider a solid torus tube $H$ around a disclination $K$---we interpret the interior of $H$ as the core of the defect. Our goal here is to determine the magnitude of the defect charge `locally', by looking only at the dividing curve on the boundary of $H$ and using it to infer the dividing curve on a sphere $S$ enclosing the defect, analogously to how we determine the charge of a point defect by looking at the sphere that encircles the defect core. The signs of the defect charges are then determined by treating the disclinations as point defects inside $S$ and arguing as described above. 

Topologically, we can view $H$ as a cylinder with a smaller cylinder cut out, as illustrated in Fig.~\ref{fig8}. Filling this hole in with a `plug', a cylinder $C$ which is topologically a ball, will result in another topological ball $B$ which contains the disclination. Since we assume $K$ is not linked with other disclinations (it actually suffices for it to be linked with an even number of other disclinations) it follows that the director can (up to a homotopy near the boundary of $B$) be oriented over the boundary of $B$, and this can allow us to determine the defect charge of $K$ up to a sign choice. In order for this process to make sense, the director field on the plug $C$ must join smoothly to the director field in $H$ to ensure that no defects. This is where the dividing curves come in. The plug $C$ must be assigned a director field with a dividing curves so that the dividing curves line up along the hole in $H$. This ensures that the director fields agree across the join, up to a homotopy~\cite{geiges_introduction_2008}. 

The ball $B$ then contains two features: the original disclination $K$, and possibly a defect in $C$ (without loss of generality all singular behaviour in $C$ can be contained in a single point). Both of these balls $B, C$ are assumed to be oriented the same way, either both with an outwards-facing normal or both with an inwards-facing normal. If we write $q_\text{ball}$ of the defect charge in $B$ and $q_\text{plug}$ for the defect charge in $C$, then by Poincar\'e--Hopf the charge $q_\text{dscln}$ carried by $K$ must satisfy 
\begin{equation}
    q_\text{plug}+q_\text{dscln}=q_\text{ball}.
\end{equation}
This construction is especially useful in determining when a disclination has vanishing defect charge, because this does not require any choices of orientations. For example, in Fig.~\ref{fig8}(a) we visualise this construction for a defect belonging to J\"anich class $(0,0)$ with self-linking $-2$. The plug in this case must have a dividing curve with a single component, which implies it contains zero defect charge. The result of gluing in the plug is a ball whose boundary also has a dividing curve with a single component, implying that the disclination has vanishing defect charge, as we expect, and also that this is completely unambiguous. This particular type of defect loop is exactly the `figure eight' loop illustrated in Fig.~\ref{fig7}(c). 

In Fig.~\ref{fig8}(b) we illustrate the construction for a $+1/2$ winding disclination in J\"anich class $(0,1)$. The plug is a hedgehog defect, and our calculation shows that the disclination has charge $\pm 1$ and that the sign must be chosen unambiguously so as to give the opposite charge as the hedgehog. In Fig.~\ref{fig8}(c) and (d) we show the construction again, but after the $(0,1)$ line has undergone the two fundamental homotopies of its local structure, respectively a protrusion growth and a Hopf process. 

We can exploit the freedom to move components of the dividing curve around to give a simpler construction, which can be seen as shrinking down any loop in the class $(0,1)$ to a Morse point defect. It is always possible, by moving the components of the dividing curve around so that some of them sit `inside' the central cylinder of the boundary torii shown in Fig.~\ref{fig8}, to ensure that the plug has charge zero. For example, in Fig.~\ref{fig8}(c) we could move one of the nullhomotopic components of the dividing curve inside the central hole. One may do this while {\it simultaneously} ensuring that the ball that results from gluing in the plug has a dividing curve with either no components, or two components, so that it corresponds to a Morse defect. We can flip between a hedgehog structure (empty dividing curve) and hyperbolic structure (two components to dividing curve) by performing a homotopy of the disclination line using the protrusion growth process, a disclination-line version of the local homotopy between the corresponding Morse defects illustrated in Fig.~\ref{fig1}(b).

Thus, as long as all disclination lines are unlinked from one another it is possible to surround each individual feature with a ball so that the interior is topologically equivalent to a Morse defect; this means that each disclination line can be shrunk to an effective point defect, with Morse behaviour. Once this assignment is made, the ball can be assigned a Morse Index and hence a defect charge by orienting the director across it as described above and illustrated in Fig.~\ref{fig7}, and this assignment is then well-defined up to a global flip. For links, we may use a surface of higher genus (e.g., a genus 2 handlebody for a Hopf link) and perform the same construction in the more general setting, with $g$ plugs. 

The charge defined in this way is then entirely unambiguous, locally consistent, and globally obeys the Poincar\'e--Hopf theorem. It is invariant under all homotopies of the texture, even if they chance the local structure of the disclination, so long as those homotopies do not involve the merging or splitting of defects or rewiring processes---it is additive under latter. It does not require a constituent global orientation of the director, and does not induce or require a consistent orientation for the far-field; it requires only the orientation of a finite set of integral curves connecting the defects.

\section{The Topological Classification of Rewiring Processes}
\label{sec:rewiring}
Rewirings are a fundamental process by which segments of disclination lines merge and split, and are the basic means by which disclination lines interact. They have been discussed extensively in the context of defects forming around colloidal particles, where the basic textures can be related to each other by these rewirings~\cite{tkalec_reconfigurable_2011, copar_PRE_2011, copar_nematic_2011, copar_elementary_2013, copar_topology_2014, copar_knot_2015, musevic_liquid_2017}, and also in active nematics~\cite{copar_topology_2019}. In this section we give a full topological classification of rewiring processes in terms of tomography and singularity theory. This classification reveals that there are in fact several different classes of rewiring which are geometrically and topologically distinct, and also clearly elucidates the role merons play in mediating these interactions. 

\begin{figure}[t]
\centering
\includegraphics[width=0.98\linewidth]{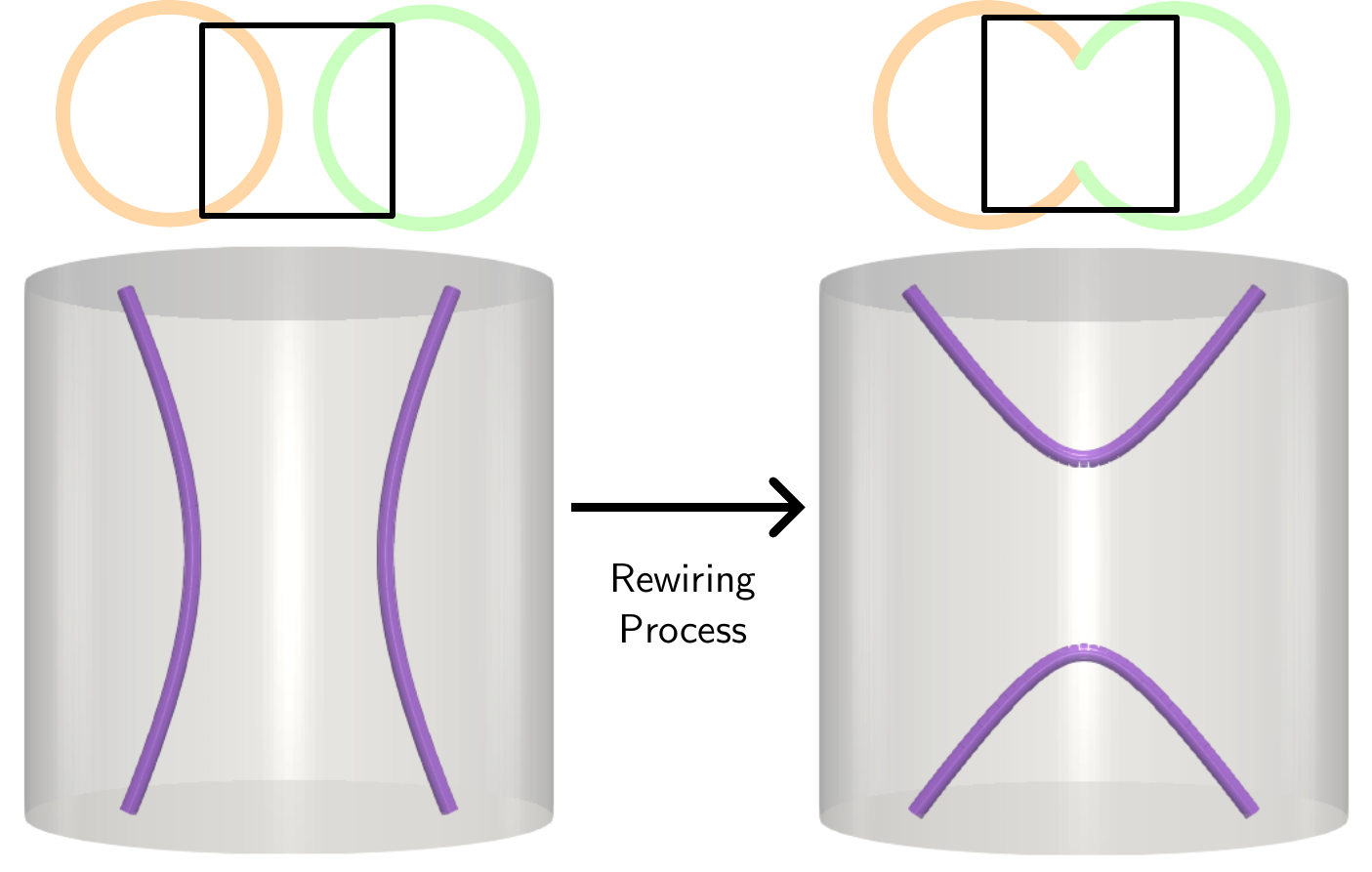}
\caption{A rewiring process involves taking two segments of disclination lines (purple) inside a cylinder (left), and merging them in the core of the cylinder in order to change the connectivity (right). This process can be viewed as a surgery, by excising the cylinder on the left from the material and replacing it with the texture in the cylinder on the right, or as a dynamical process that occurs in the material.}
\label{fig9}
\end{figure}

Formally, we define a rewiring as a process involving a topological ball $B$ (which we picture as a cylinder, as in the previous section) inside the material. We imagine that $B$ intersects two strands of disclination line, as illustrated in Fig.~\ref{fig9}---these may come from two different disclinations, or be two segments of the same disclination. A rewiring involves either (i) a surgery that removes this ball followed by gluing in a replacement which changes the connectivity of the disclination strands, with the dividing curves on the boundary lining up with those on the original ball, or (ii) a dynamical process in which two strands merge in the middle and then separate in an orthogonal direction to change the connectivity. We adopt the more physical perspective of rewiring as a dynamical process that occurs over some normalised time $t \in [0,1]$. We can find local coordinates $x,y,z$ in $B$---for our models we assume that $x,y$ satisfy $x^2+y^2 = 4$ and $z \in [-\pi, \pi]$---and perform a tomography using disks of constant $z$ value. Locally, the director can be made tangent to all of these disks at the start of the rewiring process---though it may not remain so if the rewiring involves a meron---and thus can be described by a 2-parameter family ${\bf q}_{z,t}$ of 2D vector fields, packaged into a $t$-parameterised family of Q-tensors as described in Section \ref{sec:tomography}. 

\begin{figure*}
\centering
\includegraphics[width=0.98\textwidth]{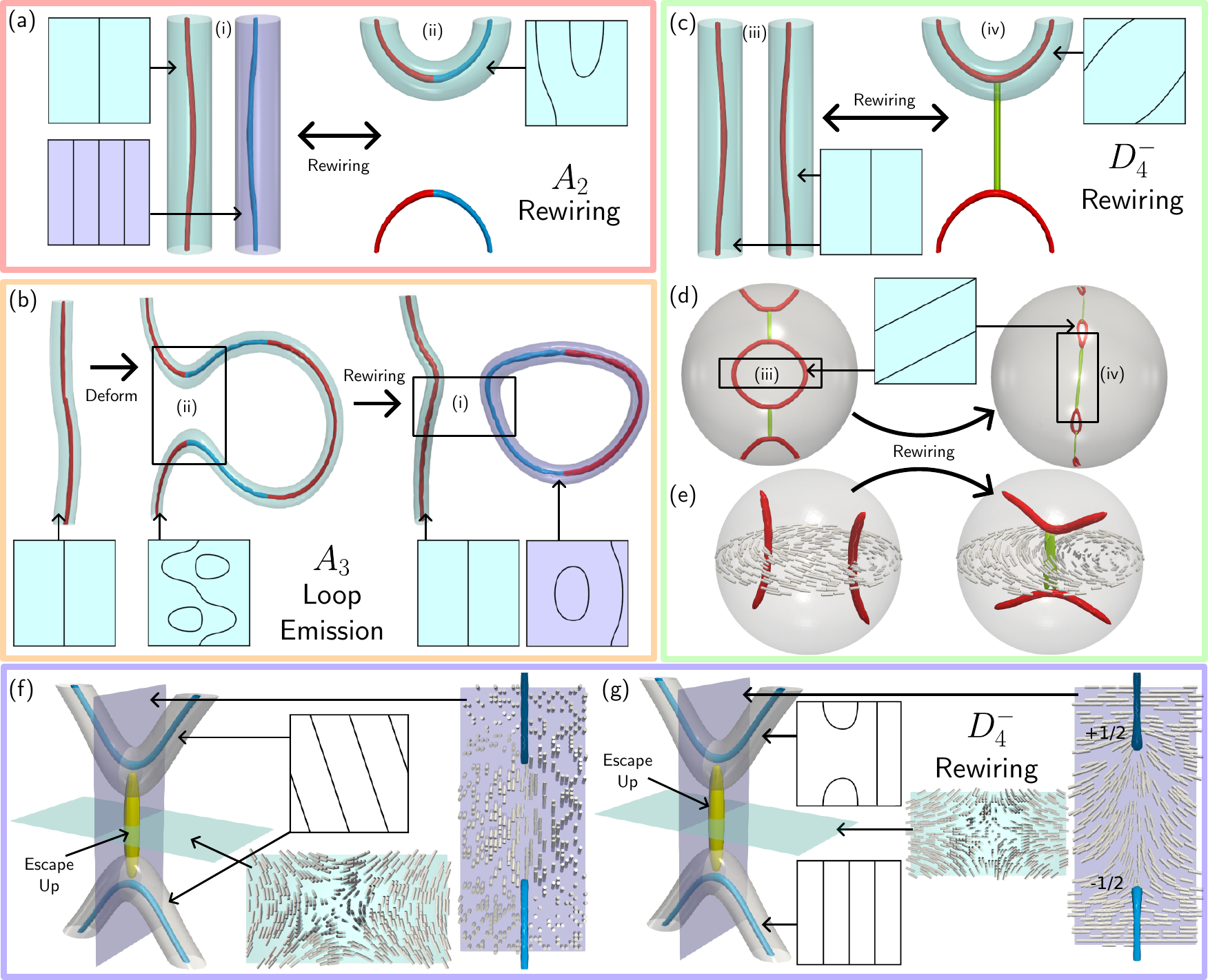}
\caption{Rewiring processes are the fundamental means by which disclination lines interact with one another, analogous to (but also far more complicated than) the creation and annihilation events of point defects. The result of a rewiring is to either introduce or remove structural degeneracies in the tomography, and the processes can be classified topologically by examining these degeneracies. (a) Rewirings of $A_2$ type involve strands of opposite winding. (i) We begin with two segments of disclination line, along which we may `straighten' the dividing curve, as winding is a global property. A rewiring process as described in the main text produces the picture (ii), and changes the dividing curve on each segment as illustrated. (b) This process occurs during the emission of a defect loop from another disclination. The disclination first undergoes a protrusion growth homotopy. This results in a region that is topologically the same as that shown in (a)(ii). Rewiring then replaces this with the picture in (a)(i), and the result is the separating off of a loop. The dividing curves are as shown; the loop is always of J\"anich class $(0,0)$. This entire spatio-temporal process can be described by tomography and an unfolding of the $A_3$ singularity. (c) $D_4^-$ rewirings involve strands of the same winding; we illustrate $+1/2$, but the $-1/2$ case is analogous. These rewirings always result in a meron line. The winding of this depends on the exact profile and is important, as described in the main text---we illustrate \eqref{eq:plus1_rotational}. (d) A texture in a droplet of active cholesteric with planar anchoring, observed in the numerical simulations of Ref.~\cite{carenza_chaotic_2020}. The rewiring process describes the disintegration of a defect loop of J\"anich class $(0,0)$ and self-linking $+2$ into two loops of the same class and winding. (e) Another dynamical texture observed in the numerical simulations of Ref.~\cite{carenza_rotation_2019} involves two disclination lines that `dance'---this is nothing more than another realisation of the rewiring process. (f,g) For each winding number $\pm 1/2$ The are two distinct versions of the $D_4^-$ rewiring, which we illustrate here for the case of $-1/2$ winding. These depends on the relative orientations of the singularities in the cross-section. (f) The $-1/2$ analogue of the process shown in panels (c-e) puts a Dehn twist into the dividing curves of the lines after the rewiring process---this is given by~\eqref{eq:rewiring_minus_a}. The director field is shown on two cross-sections (pale blue, purple), and we also show the dividing curve on two tubes around the disclinations (white). There is a $-1$-winding meron (yellow). (g) An alternative process, involving a $\pi/2$ rotation of each director stick, results in a different outcome that does not introduce winding to the profile of the dividing curve---this is given by~\eqref{eq:rewiring_minus_b}. While the disclinations induce a $-1/2$ winding singularity into each $z$-surface, the projection into an $x$-surface (orthogonal to the disclination) shows a $+1/2$ singularity in the winding of the projection. For $+1/2$ winding the distinction between these is the distinction between a rotational and radial meron, as described in the main text.}
\label{fig10}
\end{figure*}

There are two classes of rewiring: one which involves strands of the opposite winding and is described by the $A_2$ singularity classes, and another which involves strands of the same winding and is described by the $D_4^-$ singularity classes---the fundamental distinction is then between the types of degeneracies which appear in the tomography and whether or not these involve merons, see Fig.~\ref{fig3}. Within these classes are distinct forms which depend on the relative orientations of the singularities in the cross-sections, analogous to what we have already described in Section~\ref{sec:tomography} regarding the local structure of degeneracies, and in Section~\ref{sec:self_linking} regarding changes to class $(0,0)$ disclination loops. 

\subsection{Rewiring of Strands of Opposite Winding}
First we describe the rewiring of strands of the opposite winding. This is illustrated in Fig.~\ref{fig10}(a). Up to homotopy the dividing curves on the strands are straight lines, panel (i). Once we rewire, panel (ii), the disclination segments are half $+1/2$ profile and half $-1/2$ profile, so we have replaced two non-degenerate segments of disclination with a pair of fold degeneracies. The dividing curve is as indicated in the insets of Fig.~\ref{fig10}(a). There are two geometrically distinct types of the 2D form of the $A_2$ singularity: one where it splits into a pair of Morse singularities with MI 0 and MI 1, and another type where it splits into a pair of Morse singularities with MI 1 and MI 2. The distinction between the two cases is purely about whether the longitudinal component of the dividing curve veers to the left, or to the right, of the other component---we show the left-veering case in Fig.~\ref{fig10}(a)(ii). 

We give models for these rewirings based on unfoldings of the $A_2$ singularity. As we increase the time parameter $t \in [0,1]$, the disclination strands will merge in the the region $z \in [-\pi/2, \pi,2]$. The two types of rewiring process, labelled $A_2^{\pm}$, can be described using tomography via the the following two-parameter (space $z$ and time $t$) families of vector fields, which can be identified with two-parameter families of Q-tensors as described in Section \ref{sec:tomography}:
\begin{equation}
    {\bf q}^{\pm \pm}_{A_2} = \pm \left(x^2 -1 + t(1-\cos z ) \right){\bf e}_x \pm y {\bf e}_y.
\end{equation}
Making opposite choices of the signs simply swaps the position of the $+1/2$ and $-1/2$ singularities, while making the same choice rotates the profiles, switching between the case where the singularities have MI 0 and MI 1, and the case where they have MI 1 and MI 2. By means of a different model for an $A_2$ unfolding, for example described in Refs.~\cite{tang_orientation_2017, tang_theory_2019}, we may rotate the profiles of the defect strands independently of one another; while this alters the energetics it does not effect the topological aspects of the process. 

This type of rewiring either introduces or destroys pairs of $A_2$ `fold' degeneracies in the tomography, and therefore does not involve mediation by merons. Played out in reverse, the rewiring is part of the process by which a disclination line emits a loop. This process is illustrated in Fig.~\ref{fig10}(b). Firstly a disclination segment undergoes a protrusion growth process, illustrated in Fig.~\ref{fig4}(e,g). This introduces four `fold' degeneracies into the tomography. Then a rewiring takes place to pinch off the protrusion, removing two of these fold degeneracies; the other two remain as part of the loop that has been emitted. Alternatively, this entire process can be described via tomography using a two-parameter unfolding of the $A_3$ singularity to capture both the protrusion growth and the rewiring.

\subsection{Rewiring of Strands of Like Winding}
The topological and geometric distinction between the different rewirings corresponding to $D_4^-$ is more fundamental than the distinction between the $A_2$ types. The $D_4^-$ singularity has winding $\pm 2$ according to the choice of signs, which correspond to rewirings of $\pm 1/2$ strands respectively. For each winding, there are two distinct cases. 

We consider $+1/2$ first. The two $+1/2$ singularities must have distinct Morse Indices 0 and 2. The rewiring merges them in the middle, creating a $+1$-winding singular line between the rewired sections. There is a geometric distinction according to whether the `heads' of the $+1/2$ singularities face each other or face away from each other. In the former case, the $+1$ singularity that stretches between the two arcs resulting from the rewiring is splay-type. This is described by the toy model,
\begin{equation}
    {\bf q}_{D_4^-}^{-+} = -\left( y^2-x^2 -f(t,z) \right)\,{\bf e}_x + 2xy\,{\bf e}_y,
\end{equation}
where the unfolding is controlled by the function,
\begin{equation}
    f(t,z) = \begin{cases}
                  \cos z   & |z| > \pi/2,\\
                  0 & \text{otherwise}.
             \end{cases}
\end{equation}
Between $z=-\pi/2, z=\pi/2$ at $t=1$ there is a $+1$-winding singular line which  may either escape up or escape down, regardless of material chirality~\cite{pollard_escape_2024}. The result is to introduce a nullhomotopic loop into the dividing curve of one of the line segments. This rewiring results in the creation of two $D_4^-$ `elliptic umbilic' degeneracies. We do not illustrate this here, but do so presently for the equivalent $k=-1/2$ structure. The other case is described by the vector field
\begin{equation} \label{eq:plus1_rotational}
    {\bf q}_{D_4^-}^{+-} = \left(y^2-x^2 -f(t,z) \right)\,{\bf e}_x -2xy\, {\bf e}_y
\end{equation}
Between the surfaces $z=-\pi/2, z=\pi/2$ at $t=1$ there is a $+1$ singular line with a rotational structure. This may either escape up, or down; the former case is left-handed and changes the self-linking of each component by $-1$, the latter is right-handed and changes the self-linking by $+1$~\cite{pollard_escape_2024}. This is the case illustrated in Fig.~\ref{fig10}(c), and we see it again involves the creation of a pair of $D_4^-$ `elliptic umbilic' degeneracies. 

We give two examples of this rewiring which have been observed in simulations of active chiral nematics in a spherical droplet with planar anchoring~\cite{carenza_rotation_2019, carenza_chaotic_2020}, Fig.~\ref{fig10}(d,e). In one case, Fig.~\ref{fig10}(d), there exist `boojum' defects at the poles of the sphere and a disclination ring in the interior. The ring is of class J\"anich class $(0,0)$ with self-linking $+2$, and is tethered to the boojums by meron lines, which are topologically necessary as the projection of the direction into any cross-sectional disk has winding $+1$ due to the boundary condition; the degeneracies here are both of elliptic umbilic type, and therefore this structure is somewhat different from the analogous loops shown in Fig.~\ref{fig6}(c), where the degeneracies are of hyperbolic umbilic type. Over time, this loop undergoes a rewiring and splits into two loops of the same type with the same self-linking, which move towards---and are eventually absorbed into---the boojums. 

At higher activity, we observe a texture containing two $+1/2$-winding strands tethered to the boundary. These strands `dance' with one another, undergoing a series of rewiring process that transition back and forth between Fig.~\ref{fig10}(c) panels (iii) and (iv). In both cases we perform our simulations starting from an initially random director, using the same methodology and parameter values as in Refs.~\cite{carenza_rotation_2019, carenza_chaotic_2020}. 

We also give the models for $-1/2$ winding. There are again two cases, distinguished by the geometry of the profile and the effect on the dividing curve. We illustrate them both in Fig.~\ref{fig10}(f,g). One case is,
\begin{equation} \label{eq:rewiring_minus_a}
    {\bf q}^{++}_{D_4^-} = \left(y^2-x^2-f(z,t) \right){\bf e}_x+2xy\,{\bf e}_y,
\end{equation}
Between the surfaces $z=-\pi/2, z=\pi/2$ at $t=1$ there is a $-1$-winding singular line which may either escape up or down. These two cases correspond respectively to a left- and right-handed Dehn twist in the dividing curve, changing the self-linking number of the lines by $\pm 1$. Note that the distinction between a dividing curve that bends to the left and one that bends to the right is not dependent on the orientation of the disclination line itself (although the self-linking number is); rather, they are distinguished by the handedness of the Dehn twist. A numerical simulation of this case is shown in Fig.~\ref{fig10}(f) for a left-veering dividing curve, with the director shown on a slice across the meron (light blue) and also on a slice that is orthogonal to the `cusp' point of the disclinations (purple). 

The final case is 
\begin{equation} \label{eq:rewiring_minus_b}
    {\bf q}_{D_4^-}^{--} = -\left(y^2-x^2-f(z,t)\right){\bf e}_x-2xy\,{\bf e}_y,
\end{equation}
This is shown in Fig.~\ref{fig10}(g). This case is especially interesting, as it illustrates the distinction between the winding around the projection of the director field into a tomography and the winding around orthogonal disks, as seen by the dividing curve on a tube surrounding the disclination. For this case, the profile always has a total $-1$ winding, given by the meron in the central part (light blue surface in Fig.~\ref{fig10}(g)) and splitting into two $-1/2$ disclination strands. However, if we take a cross-section which is orthogonal to the disclination line at the cusp point (purple surface in Fig.~\ref{fig10}(g)), we see it has $+1/2$ winding. Indeed, we may also see this from the dividing curve, which contains a nullhomotopic component and clearly shows a region of $+1/2$ profile. This example illustrates why the tomography perspective on a material should always be supplemented by the surgery perspective of the local structure of a disclination in order to obtain a full picture of the material structure---a disparity between the two relates to the actual geometry of the line, and will be discussed in future work. 

Thus, to summarise, there are four different types of rewiring involving strands of opposite winding, two for $-1/2$ profile and two for $+1/2$ profile. Each case corresponds to a different (nondiffeomopric) form of the $D_4^-$ singularity. For each kind of winding, one of the cases results in one of the lines picking up a small region of the opposite winding, with no change in the self-linking number. The other case has two subcases, corresponding to a left- or right-handed Dehn twist in the dividing curve; these subcases are energetically equivalent in an achiral material but in a chiral material there may be a preference depending on the material's handedness~\cite{pollard_escape_2024}. 

These distinctions may manifest themselves in the Frank--Read mechanism in a liquid crystal~\cite{long_frank-read_2024}. A Frank--Read source corresponds to a disclination arc whose endpoints both lie on a boundary component. Such an arc can act as a source for the emission of disclination loops by a series of rewiring processes. This mechanism can be induced by passive elasticity or by active forces~\cite{long_frank-read_2024}. A Frank--Read source is `half' of a disclination loop, attached to the boundary of the material. It can be comprised of a $+1/2$ and a $-1/2$ segment in the form of half of a $(0,0)$-type loop---we call this an $A_2$ source---or it can have a constant $+1/2$ or $-1/2$ profile, being half of a $(0,1)$-type loop---a $D_4^-$ source. The emitted loops have distinct J\"anich invariants, linking numbers, charge, and very different geometric structure. 

The $A_2$-type Frank--Read source emits $(0,0)$ loops with no self-linking. These can either have the longitudinal component on the `inside' (a loop with `radial twist') or the `outside' (a loop with `tangential twist') according to which form the $A_2$ singularity the source corresponds to. The $D_4^-$-type sources can emit loops of class $(0,0)$ with self-linking $\pm 2$ (the case illustrated in Fig.~\ref{fig8}(a)), or they can emit loops of type $(0,0)$ with no self-linking, with the distinction being down to the relative orientations of the profile as in Fig.~\ref{fig10}(f,g). In each case there is a sharp distinction from the $A_2$-type sources: the loop emitted remains tethered to the source by a meron line. The dynamical process illustrated in Fig,~\ref{fig10}(d) is analogous to the emission of a loop by a Frank--Read source---one can see that it involves twisting of the `sources' that are tethered to the boundary---except played in reverse, so that the sources absorb the loops rather than emit them.

\begin{figure*}
\centering
\includegraphics[width=0.98\textwidth]{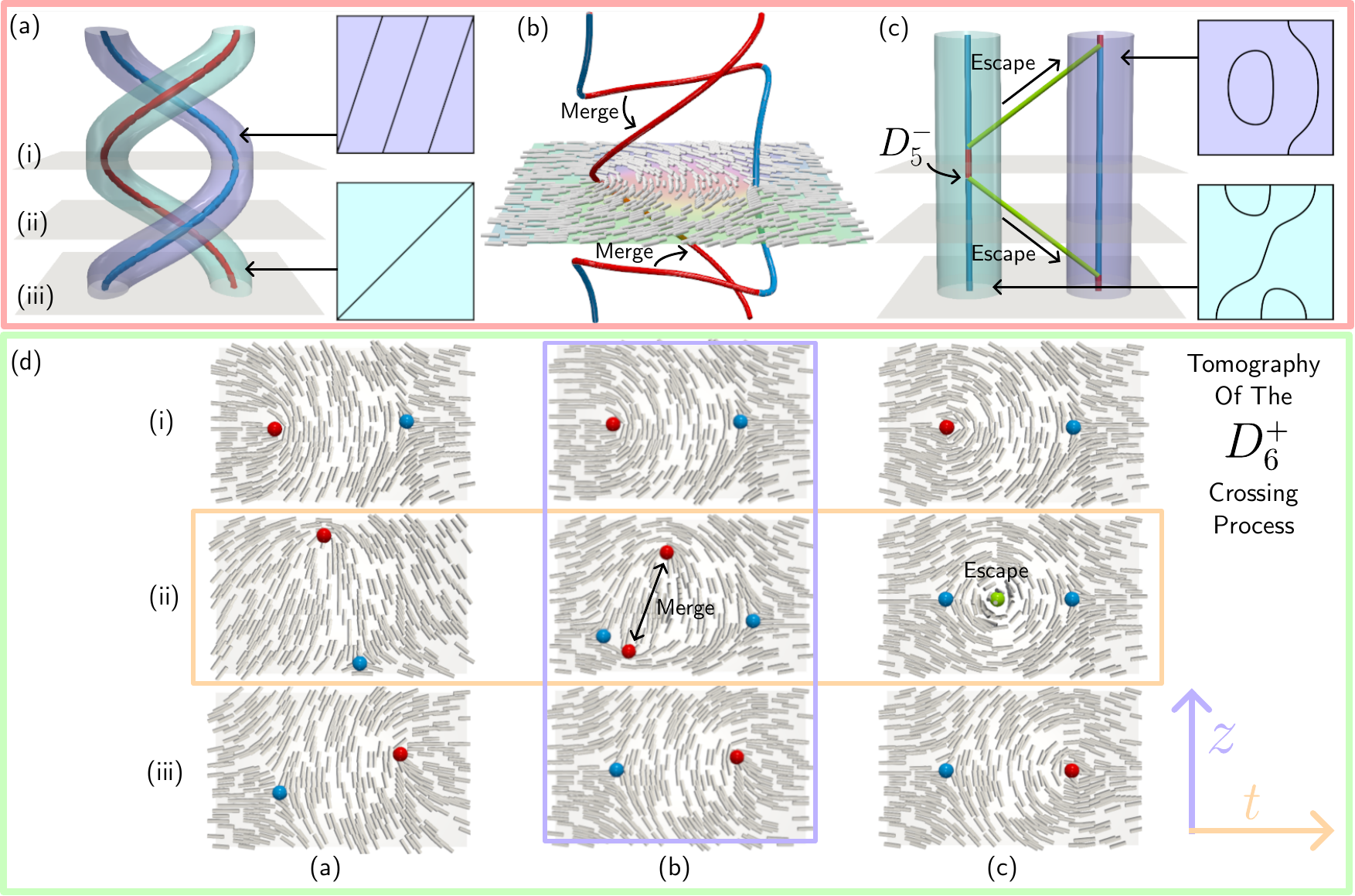}
\caption{A tomographic analysis of the crossing of disclination strands of opposite winding shows how this process results in the formation of meron tethers between the lines. (a-c) Snapshots from the crossing process described in the main text. (a) We begin with a pair of linked disclination strands with opposite winding $+1/2$ (red) and $-1/2$ (blue). (b) The crossing process proceeds by having the $-1/2$ strand deform by a protrusion growth process, resulting in the appearance of segments of $+1/2$ winding. These then merge with segments of the other disclination. (c) The merged strands are $+1$ singular lines which escape to give meron tubes (green). The disclination strands are now unlinked, but the meron tethers remain. (d) We examine this process spatio-temporally through the tomography. We illustrate the projection of the director on three surfaces, labelled (i) to (iii) in panel (a). Going across from (a) to (c), we show how the structure of the projection changes over time.}
\label{fig11}
\end{figure*}

\section{Crossing Processes}
\label{sec:crossing}
In addition to rewirings, disclination lines may cross one another, a process which necessarily results in meron lines tethering the disclinations together~\cite{poenaru_crossing_1977, alexander_entanglements_2022}. Crossings cost more energy than rewirings and are therefore much less likely to be observed, although they can be seen in the defect coarsening process that follows the isotropic-nematic transition~\cite{chuang_coarsening_1993}, and potentially in active nematics as well. Often what are called crossing processes in the literature are actually what we have called rewiring processes~\cite{ishikawa_crossing_1998}. 

For completeness, and for a nice illustration of the tomography technique, we will describe the crossing of strands of opposite windings. This is illustrated in Fig.~\ref{fig11}. It corresponds to a complicated two-parameter unfolding of the $D_6^+$ singularity that we describe presently. Snapshots from this process are shown in Fig.~\ref{fig11}(a-c). We perform a tomography on the bottom half of the structure (the top half is simply the mirror image), illustrating the director on three slices labelled (i-iii) in Fig.~\ref{fig11}(d). Part of the process is the merging of strands of the same winding, in this case $+1/2$, which then escape to the form the meron tethers. The `bending back' the occurs in one of the disclination lines, Fig.~\ref{fig11}(b) involves the protrusion growth process shown in Fig.~\ref{fig4}(e,g), and results in the creation of four $A_2$ `fold' degeneracies in the tomography. The result of merging these strands is to replace the `fold' degeneracies with $D_5^-$ `parabolic umbilic' degeneracies. We see the result of this process is, in each case, to change the self-linking of the disclinations by $\pm 1$, and add a nullhomotopic component to the dividing curve. 

We give an explicit functional form for the structure shown in Fig.~\ref{fig11}. As with rewiring processes, we describe the crossing of two strands inside a small topological ball with some local coordinates $x^2+y^2 = 4$ and $z \in [-\pi, \pi]$. The explicit functional form used to describe the structure shown in Fig. 5(c) is 
\begin{equation}
\begin{aligned}
    {\bf q}^\text{tethered} &= \left[y^2 + (x+\cos z)^2(x+1)(x-1) \right]{\bf e}_x\\
    & \ \ \ \ \ -2(x+\cos z)y \, {\bf e}_y.
\end{aligned}
\end{equation}
At $\cos z=1$ we have a singularity of winding $+1$ at $(-1,0)$ and of winding $-1$ at $(+1,0)$. These singularities remain fixed throughout. As we step through $\cos z $ from $1$ to $-1$ a new singularity of winding $+2$ is born and travels along the path $(-\cos z,0)$ until it reaches the other singularity, which this effects the changing of the winding number. These must be escaped to give the meron lines. Locally, each of the points where the merons and disclinations intersect is a parabolic umbilic degeneracy in the defect set, as shown in Fig.~\ref{fig3}(e). 

To describe a director with linked strands, define two functions by 
\begin{equation} \label{eq:rotating_coordinates}
    \begin{aligned}
        u(x,y,z) &= x\cos z + y\sin z, \\
        v(x,y,z) &= y\cos z - x\sin z.
    \end{aligned}
\end{equation}
This represents a rotating coordinate system for the planes of constant $z$. It rotates in an anticlockwise (right-handed) sense as we increase $z$; we can of course also define a coordinate system that rotates in a clockwise (left-handed) sense, simply by replacing $x$ with $-x$ or $y$ with $-y$. Now we consider the following one-parameter family of vector fields 
\begin{equation} \label{eq:a2_crossing}
    {\bf q}^\text{linked} = \left((u^2-1)(4-v^2)+v^2\right){\bf e}_x + v(4-v^2)\, {\bf e}_y.
\end{equation}
We may construct a $z$-parameterised family of Q-tensors using this vector field. The resulting texture has a pair of disclination lines at the points $u=\pm 1, v=0$, one with $+1/2$ and the other with $-1/2$ winding, and they are linked once: see Fig.~\ref{fig11}(a).

We produce the full process shown in Fig. 5(a-c) by interpolating linearly between these two vector fields, i.e. we take the two-parameter family $(1-t){\bf q}^\text{linked} + t{\bf q}^\text{tethered}$ for $t$ a normalised time parameter. This two-parameter family of vector fields should be identified with a two-parameter family of Q-tensors as described in Section \ref{sec:tomography}, which gives the tomography of the crossing process over time.

\section{Discussion}
\label{sec:discussion}
The growing interest in the complex spatio-temporal defect patterns observed in 3D active nematics drives a need for new theories to describe and explain these structures. In this work we have introduced a powerful new analytic framework for nematic director fields based on ideas from Morse theory. Our analysis is applicable to both passive and active systems and, by accounting for the extra topological structure, it can be applied to materials with broken symmetry in the ground state, such as smectics and cholesterics. For example, in a cholesteric, the energetic requirement of maintaining a consistent sense of handedness restricts us to the use of Dehn twists with the same sense of handedness. More constraints on the dividing curve in cholesterics are discussed in Refs.~\cite{geiges_introduction_2008, pollard_contact_2023, pollard_escape_2024}, and some of the restrictions that apply to smectics were described, albeit from a different perspective, in Ref.~\cite{machon_aspects_2019}. We have outlined the ways in which our method relates to experimental techniques, and also to other theoretical studies of nematic materials. Our method highlights the important, and often unappreciated, role that merons play in mediating interactions between defects in nematics. 

We have given a topological classification of the kinds of degeneracies that can occur in nematic disclination lines, points where the winding of the disclination changes sign or where it attaches to a meron line tether. The use of ideas from surgery theory greatly clarifies the homotopy classification of disclination lines and its relationship with an isotopy classification arising from the Volterra process. These ideas also shed light on the question of the defect charge in a material, helping to bridge the gap between global and local constraints on the charge. We have further classified the types of rewiring events that can occur in terms of singularity theory, illustrating that the winding and geometry of the profile has fundamental importance for the results of the rewiring process. This allows for an assignment of topological labels to structural changes in passive and active nematic materials, facilitating further analysis.  

In this work we have focused entirely on topological aspects of these changes in structure. Notably absent from our work is any detailed discussion of the geometry adopted by the disclination line, or any computation of energetics. As part of our classification we have provided model director fields based in singularity theory; these are designed to capture the topological aspects of structural changes and rewirings in the simplest possible manner, with no concern for the geometry or the energetic optimality. An analysis of disclinations in colloidal systems shows that there is a strong interplay between topology, geometry, and energetics, with topological properties of the disclination lines ({\it e.g.}, self-linking) manifesting in the twisting and writhing of the lines themselves~\cite{copar_topology_2014}. Furthermore, while the highly-symmetric singularity theory based models of defects show a close correspondence with simulation and experiment in systems that likewise exhibit a high degree of symmetry~\cite{pollard_point_2019}, in the general case, the exact form adopted by a director field around a complex defect structure can be quite subtle. Recent studies of the mechanics of disclination lines~\cite{long_geometry_2021, long_applications_2024, schimming_kinematics_2023} in terms of the Peach--Koehler force provide a promising avenue towards understanding how our topological constructions relate to the actual geometry of disclination lines. An alternative approach describes disclinations globally and directly in terms of their geometry via the solid angle function~\cite{binysh_maxwells_2018, houston_defect_2022, houston_active_2023}. The relationship between the solid angle and the topological structure of the director around the line is entirely unexplored. 

Along these lines, we remark that our work also highlights how complex spatio-temporal structure in nematic textures has features reminiscent of particle physics, where changes to objects (disclination lines) are mediated by solitons (merons). This suggests that there may exist analogies to Feynman diagrams, with the structural changes and rewiring processes being informally similar to particle creation/annihilation and scattering events. Feynman diagrams notably allow for an approximation of the energy of system: it would be interesting to examine whether a similar construction can be made for liquid crystals, allowing us to estimate the relative energies of different textures from tomography and/or surgery. There is also some interest in describing the transition pathways between stable, energy-minimising states~\cite{han_uniaxial_2022}. Using tomography and surgery theory, the transition pathways between energy-minimising states can be described, at least on a topological level, in a manner analogous to the calculus of surgery diagrams~\cite{kirby_topology_1989}. In materials such as cholesterics the additional topological structure restricts the possibilities considerably, and hence such materials provide an especially convenient setting for exploring this idea~\cite{gompf_4-manifolds_1999}. 

Another useful extension of the tomography technique would be to apply it not just to the director, but also to the velocity field in an active nematic. The velocity field is oriented and therefore does not have line defects, but it does contain meron lines, which in a fluid are vortex tubes. Tomography allows us to analyse the linking of these lines, which is related to vorticity~\cite{arnold_topological_2021}, and also the changes in linking over time. Vortex lines can be related to the topological entropy of the flow, and a study of their development and motion over time may help with our understanding of active turbulence in 3D. For a 2D active nematic, a tomography in the time domain may reveal subtle changes in the structure of the material than can be connected to dynamics.

Taken together, we argue that the extension, and further application of ideas rooted in Morse theory are likely to prove instrumental in the continued understanding of ordered fluids such as liquid crystals, and we therefore welcome further work in the area.

\begin{acknowledgements}
We thank G.P. Alexander (Warwick) and S.C. Al-Izzi (UNSW) for helpful discussions. We acknowledge funding from the EMBL Australia program and the Australian Research Council Centre of Excellence for Mathematical Analysis of Cellular Systems (CE230100001).
\end{acknowledgements}

\bibliographystyle{unsrt} 
\bibliography{references}
\end{document}